\documentclass[aps,prb,twocolumn,floats]{revtex4-2}
\usepackage{epsfig}
\usepackage{amsmath, amssymb}
\usepackage{graphicx}
\usepackage{soul}
\usepackage{braket}
\usepackage[colorlinks=true,linkcolor=blue]{hyperref}
\usepackage{subfigure}
\usepackage{color}

\newcommand{\blue}{\textcolor{blue}}

\begin{document}
%\raggedbottom
\title{Zero Energy States for Commensurate Hopping Modulation of a Generalized Su-Schrieffer-Heeger Chain in the Presence of a Domain Wall}
\author{Surajit Mandal$^{1,2}$}
\email{surajitmandalju@gmail.com}
\affiliation{$^1$Department of Physics, Jadavpur University, Kolkata - 700032, West Bengal, India\\
$^2$Department of Physics, AKPC Mahavidyalaya, Bengai, West Bengal -712611, India}

\begin{abstract}
We study the effect of domain wall (DW) on zero-energy states (ZESs) in the Su-Schrieffer-Heeger (SSH) chain. The chain features two fractional ZESs in the presence of such DW, one of which is localized at the edge and the other bound at the location of DW. This zero-energy DW state exhibits interesting modifications when hopping modulation is tuned periodically. We studied the energy spectra for commensurate frequencies $\theta=\pi,~\pi/2,~\pi/3$, and $\pi/4$. Following the recent study by the author of this paper [S. Mandal, S. Kar, \blue{Phys. Rev. B {\bf 109}, 195124 (2024)}], we showed numerically, along with physical intuition, that one ZES can bound at the DW position only for commensurate frequency $\theta=\frac{\pi}{2s+1}$ for zero or an integer $s$ values, while for $\theta=\frac{\pi}{2s}$ with nonzero or an integer $s$ value they appear only at the edges of the chain. We verify our numerical results by using exact analytical techniques. Both analyses indicate the realization of the Jackiw-Rebbi modes for our model only with $\theta=\frac{\pi}{2s+1}$. Moreover, the localization of zero-energy edge and DW states are investigated which reveals their localized (extended) nature for smaller (larger) $\Delta_{0}$ (amplitude of DW). The localization of topological DW states is suppressed as the width of DW ($\xi$) increases (typically scaled as $\sim1/\xi$) while the edge state shows an extended behavior only for the large $\xi$ limit.
%     \keywords{}
\end{abstract}
%\date
%\end{@twocolumnfalse}
%]
\maketitle
%\vskip 1 in
\section{Introduction}\label{sec1}
\begin{figure}[ht]
  \begin{center}
    \vskip -.6 in
    \begin{picture}(100,100)
     \put(-50,0){
       \includegraphics[width=0.8\linewidth, height= .85 in]{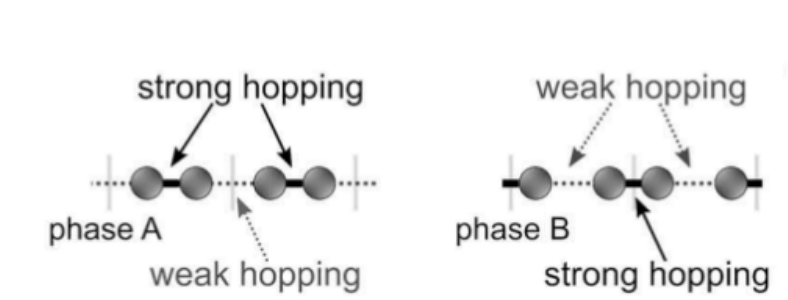}}
     \put(-60,30){(a)}
     \end{picture}
    \\\vskip -.5 in
    \begin{picture}(100,100)
      \put(-50,0){
        \includegraphics[width=0.8\linewidth,height=.85 in]{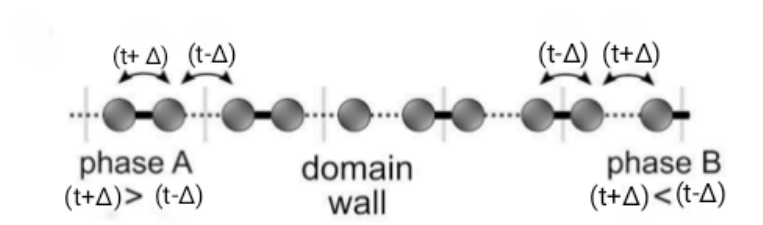}}
      \put(-60,30){(b)}
     \end{picture}
\end{center}
\caption{The two phases in the SSH model: in phase $A$, the strong hopping is inside the unit cell while in phase $B$, a strong bond exists between the unit cells. (b) Structure containing a domain wall that connects two dimerized phases of the chain and where the strong bond ($t+\Delta$) and weaker bond ($t-\Delta$) are inverted.}
\label{fig0}
\end{figure}

The Su-Schrieffer-Heeger (SSH) model, was originally introduced in the context of polyacetylene, is a one-dimensional ($1D$) tight-binding model with staggered hopping modulation where the dimerized configuration leads to two degenerate ground states\cite{su1,su2,heeger}. For a finite chain, the two staggered configurations are topologically distinct, the end site holds either a strong or a weak bond, and the specific schematic of which is shown in Fig.\ref{fig0}(a). In the topological phase, zero energy states (ZESs) appear as edge modes whose peaks die out exponentially away from the boundaries. These ZESs are sometimes familiar end solitons and carry a fractional charge $\pm e/2$ depending on the electrons added to or removed from the half-filled system\cite{rebbi,jackiw}.

The celebrated SSH model describes free fermions on a lattice with the hopping strengths switch between weak and strong
bonds and one should have a massive Dirac fermion in the continuum limit\cite{jackiw1} while in the $1D$ Jackiw-Rebbi (JR) model, Dirac fermions are generally coupled to a soliton field\cite{rebbi}. The topological defects (generally it is domain walls (DWs) in $1D$, a vortex in two dimensions, etc.) known as solitons emerge in the former cases when the arrangements of the hopping strength change from weak-strong to strong-weak at a certain lattice site, however, in the latter, soliton appear by tuning the fermion mass such that sign flips occurs at a certain point. Generally, the topological defects connect two dimerized phases (we call them phase $A$ and phase $B$) of the SSH chain creating an interface (notice schematic diagram Fig.\ref{fig0}(b)). The defects in topological systems, containing two atoms per unit cell, can host localized ZES which has fractional charges\cite{goldstone}. In the continuum limit, rather than having fractional expectation values of solitons, their charge fractionalizations can be attributed to the local charge operators representing fractional eigenvalues\cite{jackiw2}. On top of this fractionally charged background, the ZESs are familiar as excitations and are associated with MZMs\cite{scopa}. Nowadays, the exploration of exotic phases like fractional charge excitations, because of their rich fundamental physics, in modern condensed matter systems has enticed a large attraction. Such excitations are supposed to be a feasible candidate for ascertaining topological quantum computation\cite{nayak}. In particular, the fractional fermions or JR modes can be an example of such exotic excitation with
fraction charge\cite{rebbi}. The outcome of having static DWs at an interface of the SSH chain\cite{d0} and some other models\cite{d1,d2,d3} have been probed to some extent in the literature.

In this work, we keep on focusing on the periodic modulation in the hopping strength of the SSH chain including two or more than two atom unit cells. Particularly, we considered commensurate frequencies $\theta=\pi,\pi/2,\pi/3$ and $\pi/4$ for getting a periodically hopping modulated chain. Notably, the cases of $\theta=\pi,\pi/2,\pi/3$ and $\pi/4$ drive the system to become two, four, six, and eight sublattice basis respectively (discussed later). The topology and features of the edge states for different $\theta$ values have been studied to some extent in Ref.\cite{mandal}. Remarkably, the case of $\theta=\pi$ recovers the actual SSH model which, in the presence of a DW, creates fractionalized zero-energy modes that are limited in a single sublattice. In a finite chain with a DW in the form of Eq.(\ref{2}), one of these zero modes is found to appear at an edge while the other is bound to the domain wall\cite{scollon}. The intricate behavior of ZESs in the presence of such DW, however, for other commensurate $\theta$ values except $\theta=\pi$ discussed numerically to some extent in Ref.\cite{mandal}. This paper, in addition to further numerical details, provides a proper understanding of the same with more physical and analytical inspection. Specifically, we consider the JR method for the analytical study.

The paper is organized as follows. In Section \ref{sec2}, we present a brief introduction to the generalized SSH chain. The introduction of DW and the numerical spectra with DW for commensurate variation of hopping periodicity given by $\theta=\pi,~\pi/2,~\pi/3,~\pi/4$ (to be defined later) is discussed in Section \ref{sec3}. In Section \ref{sec4}, we studied numerically the behavior of ZESs for different commensurate frequencies. To concrete the numerical study, this section also includes analytical analysis. We also present the localization property of zero-energy edge and DW states in Section \ref{sec5}. Finally, in Section \ref{sec6}, we summarize and conclude our findings and discuss possible future directions and applications of work.

\section{Generalized SSH Chain}\label{sec2}
In one dimension ($1D$), the Hamiltonian for a generalized SSH chain of $L=M*N$ (in which $M$ and $N$ denote the number of sublattice and unit cells respectively) sites considering periodically modulated hopping strength can be defined as:
\begin{equation}\label{1}
  \mathcal{H}_{SSH}=\sum_{i}^{L-1}(t+\delta_{i})[\ket{i}\bra{i+1}+H.c],
\end{equation}
where $\delta_{i}=\Delta\cos[(i-1)\theta]$ with $i=1,2,3,......, n$ specifies the periodic modulation in an otherwise constant nearest-neighbor (NN) hopping strength $t$. In general for $\theta=\frac{2\pi}{M}$,  one can rewrite this as $\delta_{i+1}=\Delta\cos(2\pi i/M)$. Therefore, the system is represented by a $M\times M$ Hamiltonian matrix having $M$ number of eigenmodes. Crucially, $\theta=\pi,~\pi/2~\pi/3$ and $\pi/4$ give two, four, six, and eight sublattices in a unit cell respectively (one can see the increasing periodicity of the system for decreasing frequency). For the sake of simplicity, we consider $t=1$ as the unit of energy which further makes the unit of physical parameter dimensionless throughout this work. However, all the parameters considered here are real.
\begin{figure}[t]
   \vskip -.4 in
   \begin{picture}(100,100)
     \put(-80,0){
  \includegraphics[width=.5\linewidth]{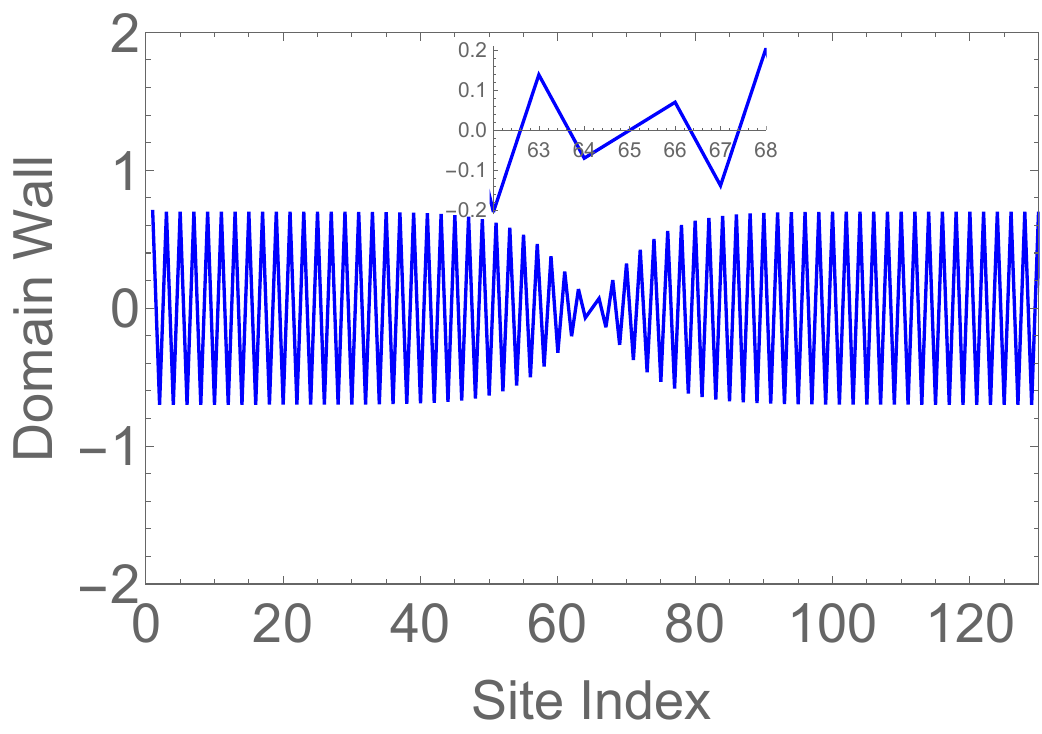}
  \includegraphics[width=.5\linewidth]{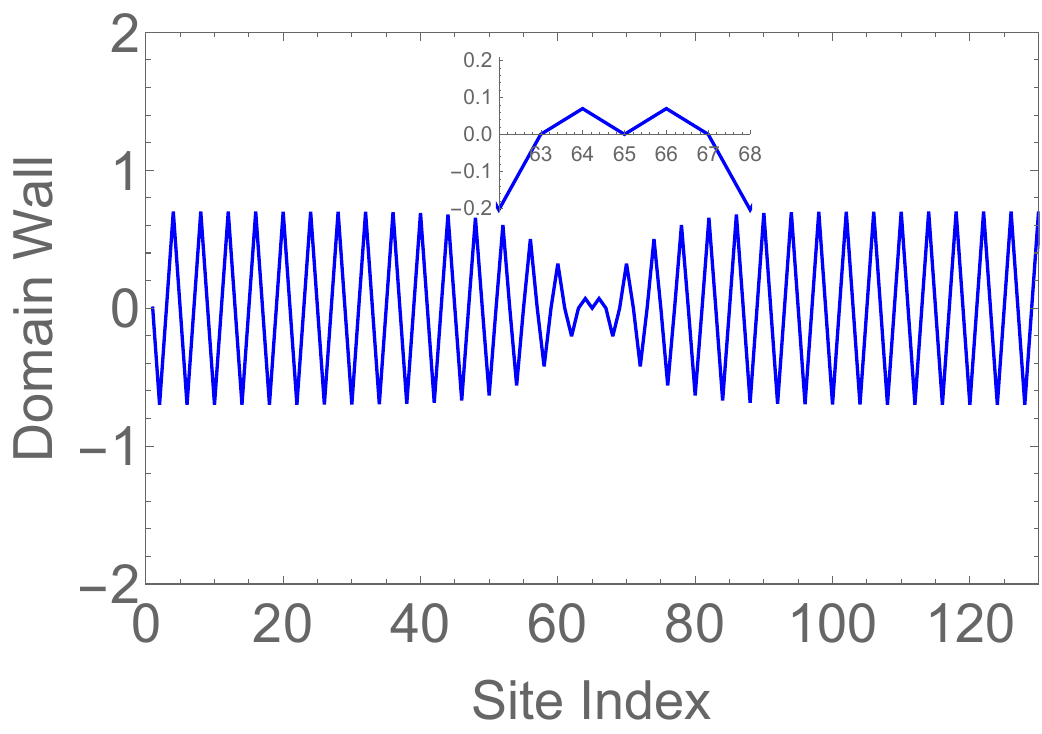}}
     \put(-49,72){(a)}
     \put(70,68){(b)}
   \end{picture}\\
   \vskip -.2 in
   \begin{picture}(100,100)
     \put(-80,0){
   \includegraphics[width=.5\linewidth]{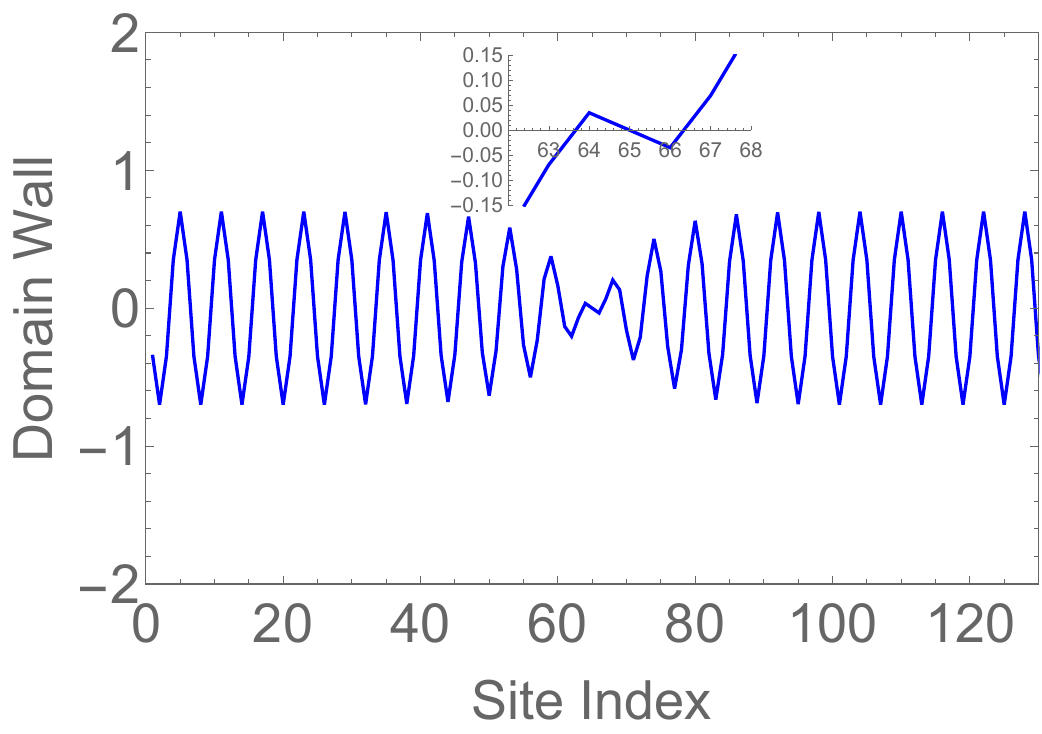}
   \includegraphics[width=.5\linewidth]{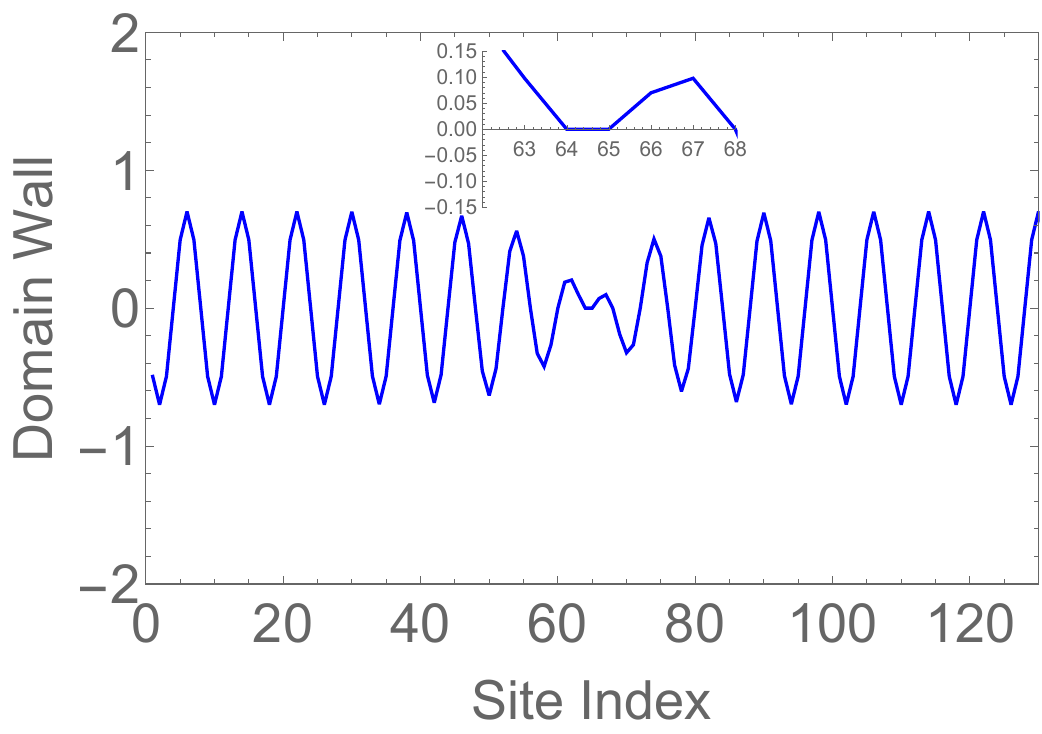}}
   \put(-49,73){(c)}
     \put(70,70){(d)}
   \end{picture}
  %\vskip -0.1 in
\caption{The behavior of kink type solitonic field for $\theta$~=~(a) $\pi$, (b) $\pi/2$, (c) $\pi/3$ and (d) $\pi/4$. The solitonic field creates a DW at $i=i_{0}$. Here, we set $i_{0}=L/2$, so the DW structure can be found at site no. $L=64$. Other choices of parameters are: $L=128$,~$\xi/a=10$,~$\Delta_{0}=0.7$. Inset shows the nature of $\delta_{DW}$ at the interface $i_{0}=L/2$.}
\label{fig1}
\end{figure}
\section{Domain wall}\label{sec3}
The engagement of a single DW in the celebrated periodically hopping modulated SSH chain boosts the domain wall (DW) to hold mid-gap ZES. This type of ZES is usually known as DW solitons\cite{mandal,kar}. In this regard, the behavior of topological DW solitons under the variation of periodic hopping modulation is of particular interest. Following\cite{mandal,scollon}, we begin by introducing a single domain wall in our model with the modulation in hopping strength now taking the following form
\begin{equation}\label{2}
\delta_{dw}=\delta_{i}\tanh\Big[\frac{i-i_{0}}{\xi/a}\Big]=\Delta_{0}\tanh\Big[\frac{i-i_{0}}{\xi/a}\Big] \cos[(i-1)\theta],
\end{equation}
in which $\Delta_{0}$, $\xi$, and $a$ denote the amplitude, the width, and the lattice spacing of the DW, respectively. Here, the index $i$ refers to the location of $i$-th site, and the parameter $i_{0}$ fixes the position of the DW. A smooth (sharp) DW can be achieved for large (small) $\xi$. This DW refers to an interface between the topological and trivial phases. It is noted that the boundaries of the SSH chain may be considered as domain walls with the vacuum or surrounding and DWs, similar to the boundary
of the SSH chain, host zero-energy localized states\cite{mandal}.

The graphical representation of $\delta_{dw}$ for commensurate frequency $\theta=\pi,~\pi/2,~\pi/3,~\pi/4$ are presented in Fig.\ref{fig1}. The plot depicts the formation of kink-shaped DWs only for $\theta=\pi$ and $\pi/3$ whereas no such DW is discernible at $\theta=\pi/2$ and $\pi/4$. The comprehensive understanding behind forming or not forming DWs at the interface is discussed more physically in the next section. Depending on the availability of DWs at the interface the ZESs decide whether they will reside in it or not. To understand this, we will present a detailed discussion both numerically and analytically in the forthcoming sections.

\begin{figure}[t]
   \vskip -.4 in
   \begin{picture}(100,100)
     \put(-80,0){
  \includegraphics[width=.5\linewidth]{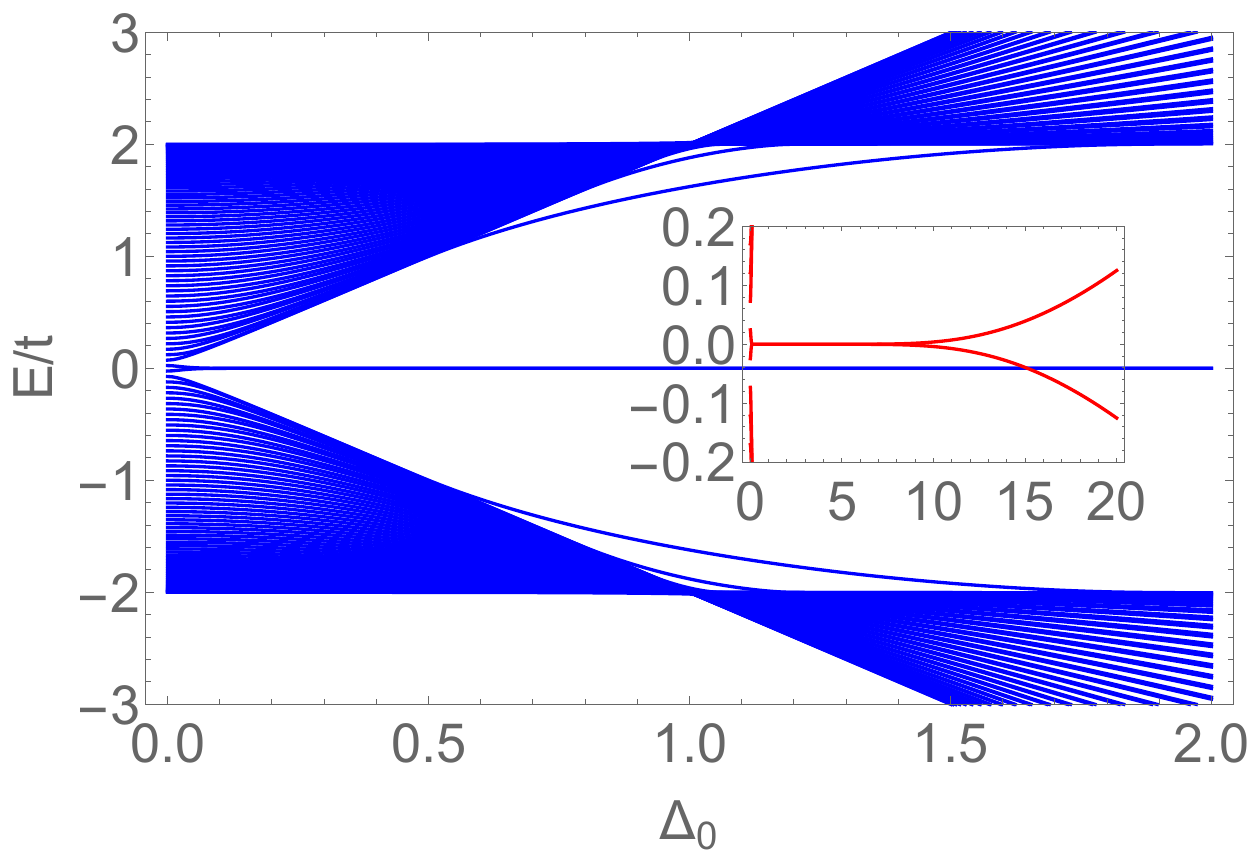}
  \includegraphics[width=.5\linewidth]{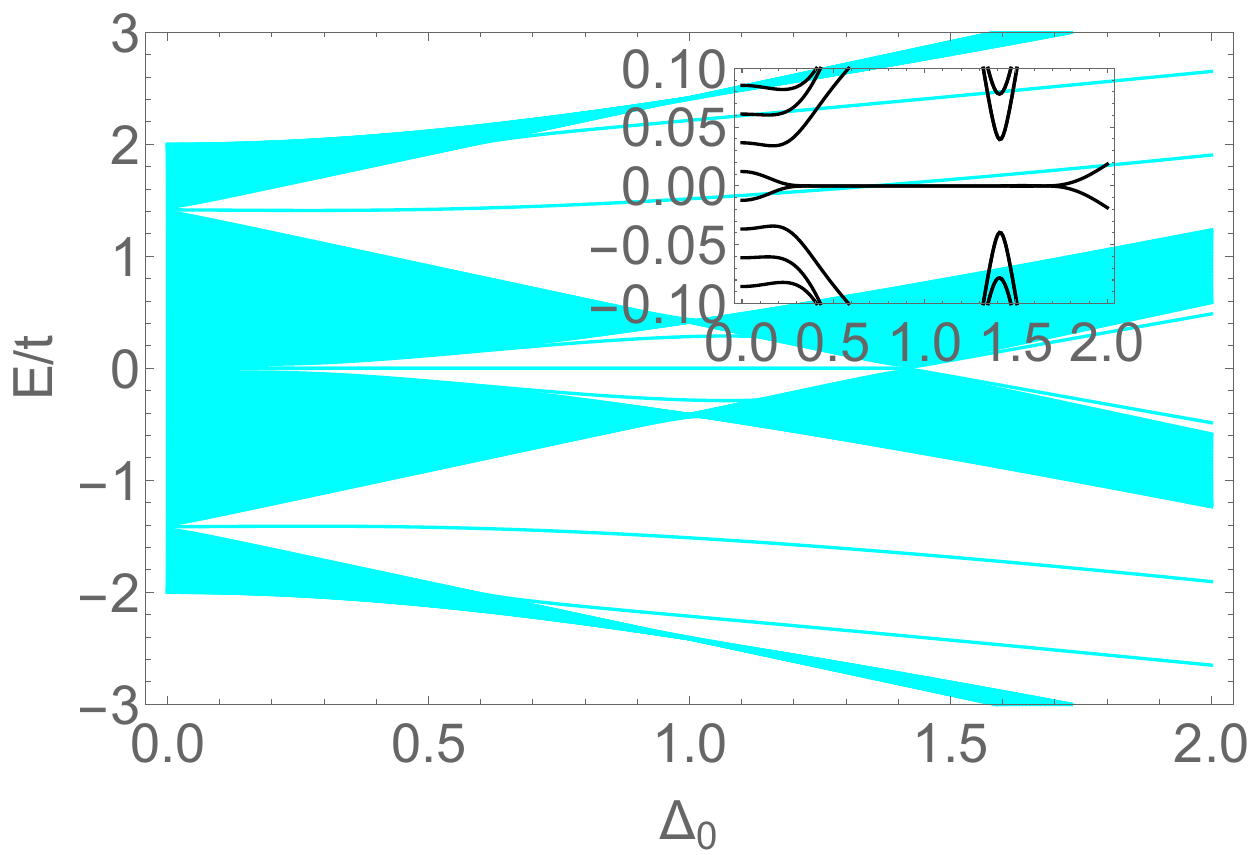}}
     \put(-49,72){(a)}
     \put(70,68){(b)}
   \end{picture}\\
   \vskip -.2 in
   \begin{picture}(100,100)
     \put(-80,0){
   \includegraphics[width=.5\linewidth]{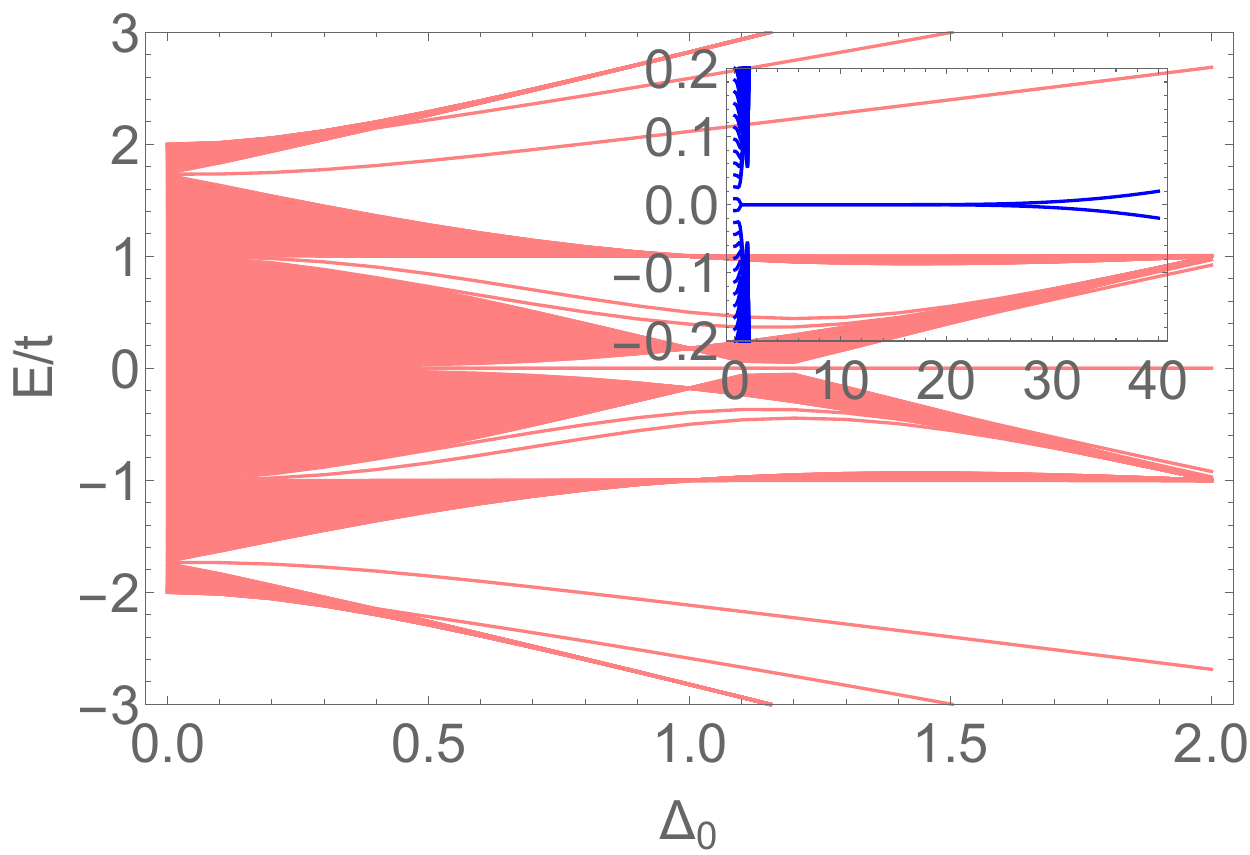}
   \includegraphics[width=.5\linewidth]{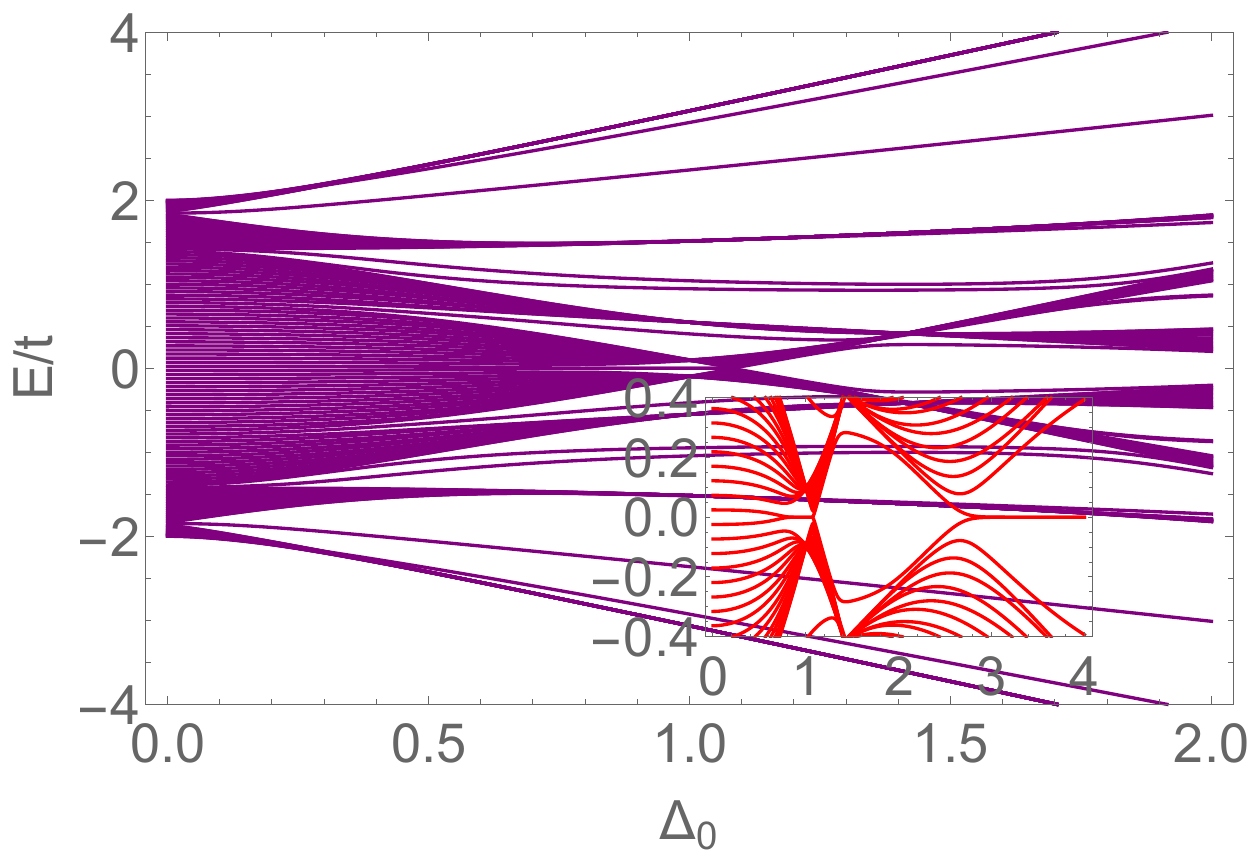}}
   \put(-49,73){(c)}
     \put(70,70){(d)}
   \end{picture}  
\caption{Numerical energy spectra with DW amplitude $\Delta_{0}$ for $\theta$~=~(a)~$\pi$ with $L=128$, (b)~$\pi/2$ with $L=256$, (c)~$\pi/3$ with $L=360$ and (d)~$\pi/4$ with $L=128$. Other choices of parameters are: ~$\xi/a=2$,~$i_{0}=L/2$. The inset shows the low energy spectra.}
\label{numerical}
\end{figure} 
\subsection{Numerical Energy Spectra}
For numerical spectra, we consider finite chains for open boundary conditions, and the chains are made out of an even number of lattice sites. The numerical energy spectra for a generalized SSH chain in the presence of a domain wall with domain wall amplitude $\Delta_{0}$ for $\theta=\pi,~\pi/2,~\pi/3$ and $\theta=\pi/4$ is reported in Fig.\ref{numerical}. We notice that the presence of DW significantly modifies the energy spectra when certain parameters are varied. In addition to the mid-gap ZESs, the DW gives rise to additional in-gap states (rather called bound states\cite{scollon}). Notably, the low energy spectra illustrate the vanishing of zero modes both for smaller and larger values of $\Delta_{0}$. One can recover a NN tight-binding model for $\Delta_{0}=0$, where no ZES is present in a finite system. For smaller $\Delta_{0}$, on the other hand, two nearly degenerate symmetric and anti-symmetric gap states subsist having almost zero energy (the energy splitting disappears for a large $L$ limit) i.e., the availability of zero modes do have a dependency on the length of the chain. For a longer chain, one can notice a smaller $\Delta_{0}$ limit that denies the existence of the ZESs which again vanishes for very large $\Delta_{0}$. In fact, the end states away from the edges decay slowly for both small and large $\Delta_{0}$ value and go as $\sim|\frac{t-\Delta_{0}}{t+\Delta_{0}}|$. These, in turn, give rise to the hybridization of the two degenerate edge modes resulting in linear combinations of symmetric and anti-symmetric states that stay (slightly) above and below the zero energy. The zero-energy bound states for this model are chiral-symmetry protected due to the presence of a pair $(E,-E)$ in the energy spectra (see Fig.\ref{numerical})\cite{AI}. The evolution of chirality-protected bound states with different kinds of disorder (such as constant, random, staggered, and interpolated on-site potentials) for various commensurate frequencies was studied elaborately in Ref.\cite{mandal}. Like $\theta=\pi,~\pi/2$ and $\pi/3$, the ZESs for $\theta=\pi/4$ will no longer persist for very small values of $\Delta_{0}$, in contrast, it will, however, be intact for very large $\Delta_{0}$. Interestingly, for intermediate $\Delta_{0}$ values, we notice the zero modes disappear within a small range of $\Delta_{0}$ values.

\section{Study of Zero energy states}\label{sec4}
In this section, we investigate the intricate behavior of mid-gap ZESs in the presence of a static DW. 
\subsection{Numerical Study}
\begin{figure}[t]
   \vskip -.4 in
   \begin{picture}(100,100)
     \put(-80,0){
  \includegraphics[width=.52\linewidth]{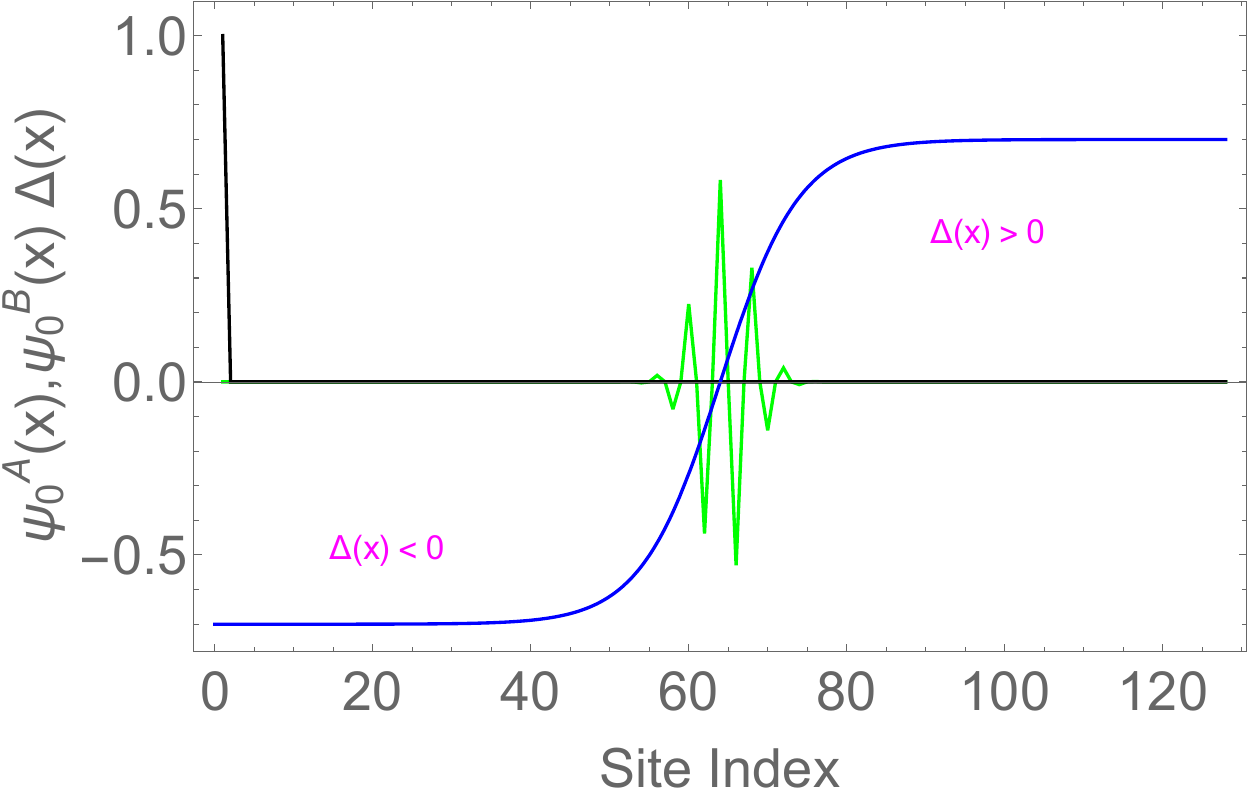}
  \includegraphics[width=.52\linewidth]{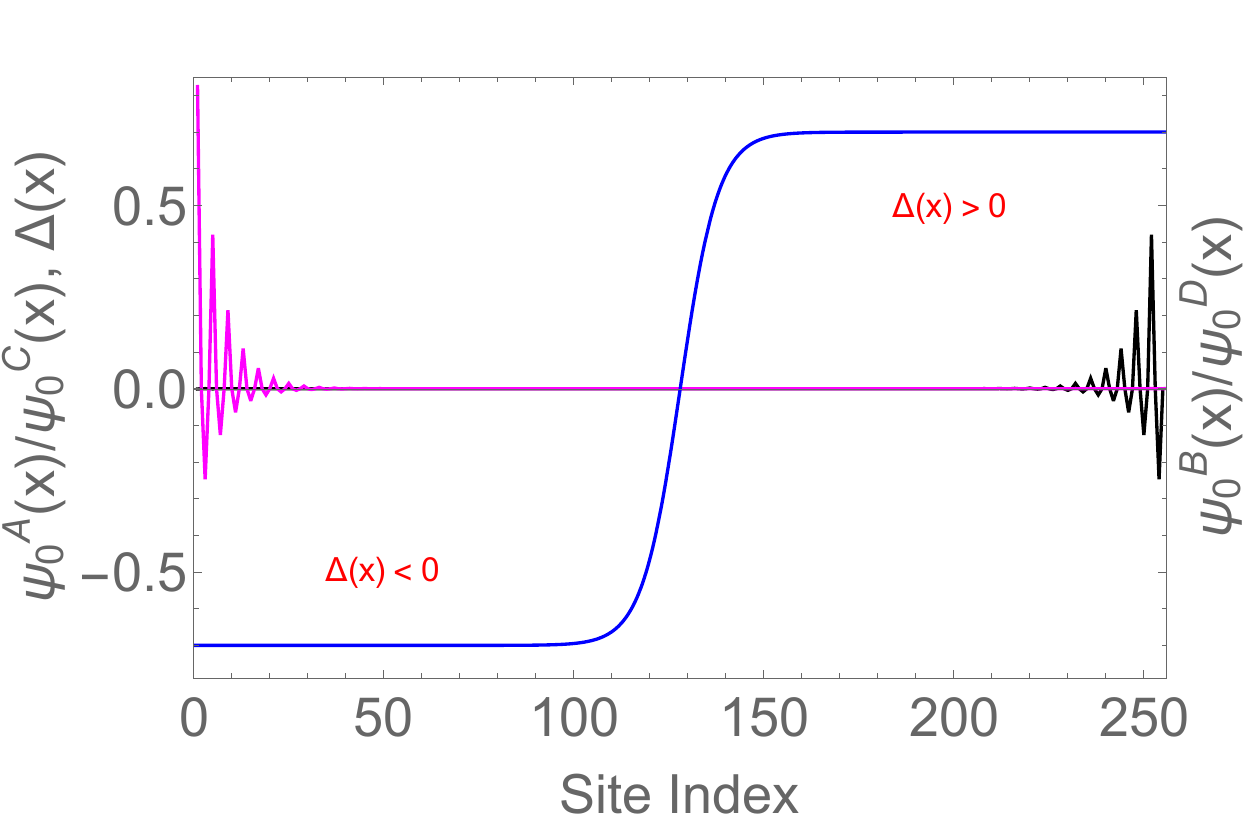}}
     \put(-49,72){(a)}
     \put(78,66){(b)}
   \end{picture}\\
   \vskip -.2 in
   \begin{picture}(100,100)
     \put(-80,0){
   \includegraphics[width=.52\linewidth]{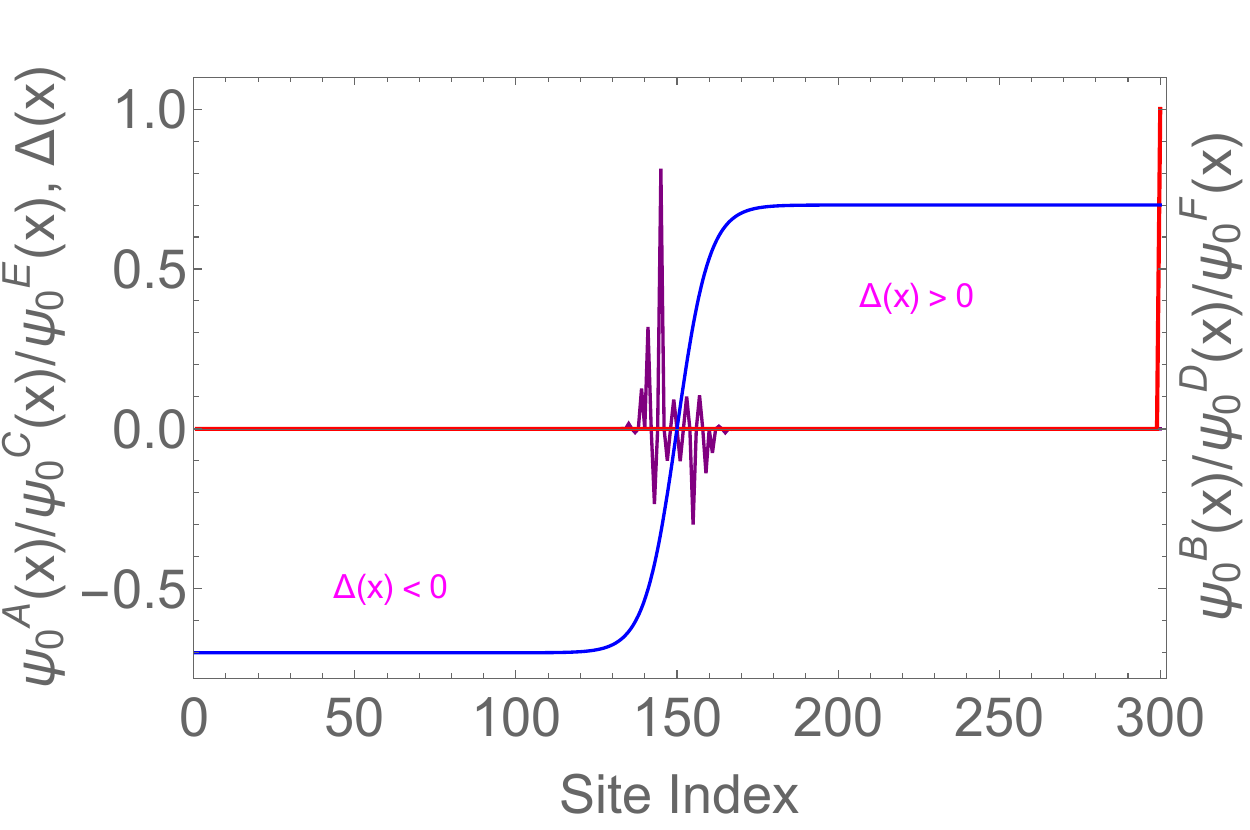}
   \includegraphics[width=.52\linewidth]{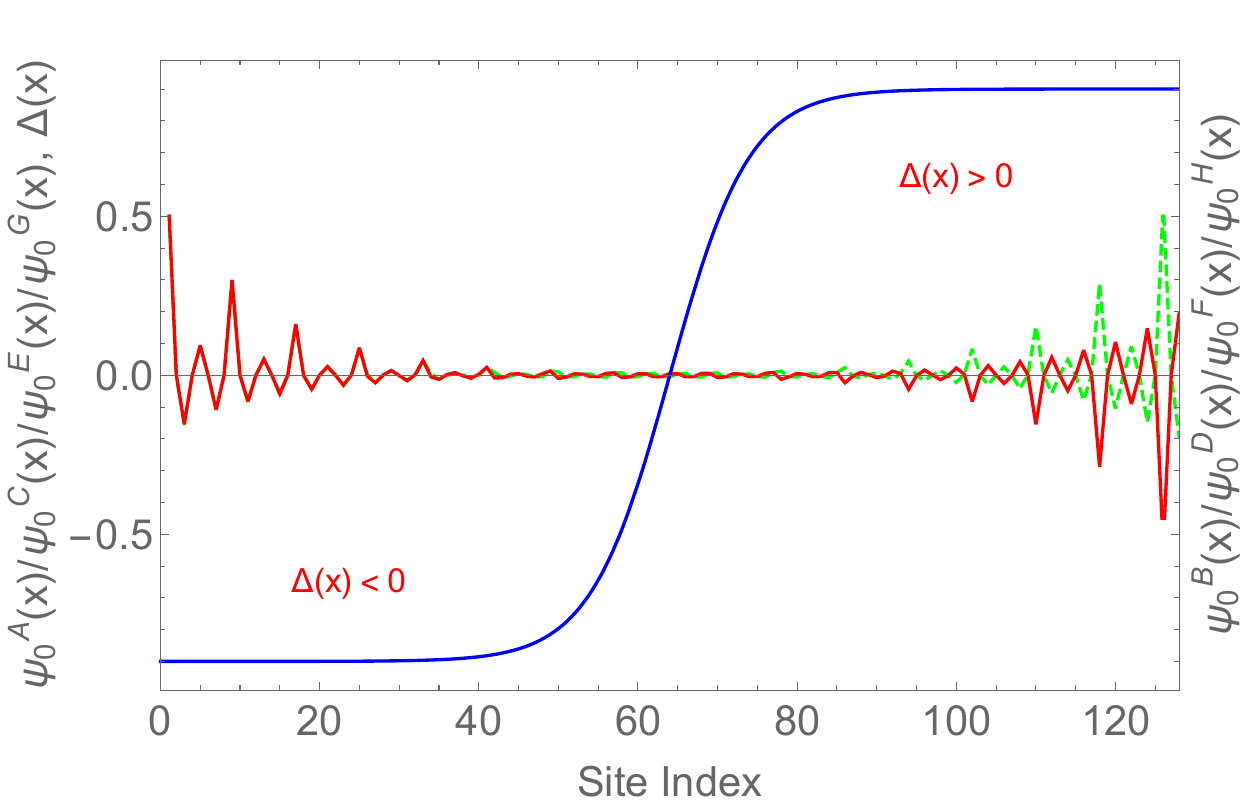}}
   \put(-49,70){(c)}
     \put(75,67){(d)}
   \end{picture}
\caption{Typical (numerical) behavior of ZESs of a finite SSH chain under consideration of DW for $\theta$~=~(a)$\pi$ where $\psi_{0}^{A}(x)$ is localized at DW position and $\psi_{0}^{B}(x)$ is localized at the left edge, (b)$\pi/2$ where $\psi_{0}^{A}(x)$ and $\psi_{0}^{C} (x)$ components are restricted at the left edge while $\psi_{0}^{B}(x)$ and $\psi_{0}^{D}(x)$ components reside on the right edge, (c) $\pi/3$ where  $\psi_{0}^{A}(x)$, $\psi_{0}^{C} (x)$ and  $\psi_{0}^{E}(x)$ components show peaks at DW while the other three at the right edge and (d)$\pi/4$ where no DW states are discernible. The kink-shaped solitonic field, $\Delta(x)$, is shown by a blue curve. The position dependent hopping strength $\Delta(x)$ interpolating between $+\Delta_{0}$ and $-\Delta_{0}$ and originates a DW at $x=x_{0}$. Here, we set $x_{0}=L/2$, so the DW structure can be found at site no. $x_{0}$. Other choices of parameters are: (a)$L=128$,~$\xi/a=10$,~$\Delta_{0}/t=1$, (b)$L=256$,~$\xi/a=10$,~$\Delta_{0}/t=0.7$, (c)$L=300$,~$\xi/a=10$,~$\Delta_{0}/t=2$, (d)$L=128$,~$\xi/a=10$,~$\Delta_{0}/t=0.9$. }
\label{zes}
\end{figure}

Now, we discuss the effect of DW on the mid-gap ZESs. Our numerical findings exhibit the localization of ZESs from one edge to the DW position and at the other edge for different commensurate frequencies. The edge states for this periodically hopping modulated SSH chain, in the absence of a DW, are localized only at the boundary of the chain\cite{mandal}. However, the typical behavior of ZESs in the presence of DW for a finite SSH chain for $\theta=\pi,~\pi/2,~\pi/3$ and $\pi/4$ is presented numerically in Fig.\ref{zes}. The ZESs are shown to be localized at the DW position only for $\theta=\pi,~\pi/3$ values while they display peaks at the boundaries for $\theta=\pi/2,~\pi/4$. Moreover, we considered the strong dimerized limit $|\Delta_{0}/t|=1$ and  $|\Delta_{0}/t|=2$ for the cases of $\theta=\pi$ and $\pi/3$ respectively which results in maximum localization of one ZES at the chain boundary [notice Fig.\ref{zes}(a)\&(c)] while the localization of zero energy DW states depends on the smoothness/sharpness of DW (rather than $\Delta_{0}$) as we discussed in Sec.\ref{sec5}.

For $\theta=\pi$, zero-energy edge states are exponentially localized in a single sublattice. Unlike $\theta=\pi$, here for $\theta=\pi/2$ the edge states are polarized in two sublattices in a way that the no two consecutive sites don't have a nonzero amplitude of them. In other words, the amplitude of the edge-state wave functions, corresponding to the two sublattices, alternatively distributed having positive and negative values while the rest of the sites (corresponding to the other two sublattices) see vanishing amplitudes in alternate sites. The tendency of alteration in the edge-state amplitude is also visible for $\theta=\pi/3$ and $\pi/4$ where the edge states are restricted in three and four sublattices respectively. The edge-states, for all the commensurate frequencies, display exponential decay of strength from the end towards the bulk of the chain\cite{mandal}. Generally, for $\theta=\pi,~\pi/3$ ZESs are restricted in an odd number of sublattices while they are restricted in an even number of sites for $\theta=\pi/2,~\pi/4$ and zero or nonzero amplitude of ZESs in alternative sites is visible for $\theta<\pi$.

From a physical viewpoint, we are introducing DW at a periodically hopping modulated parameter $\delta_{i}$ of the finite SSH chain. In the strong coupling limit $|\Delta/t|=1$, there exists an unpaired site at the DW position for $\theta=\pi$ which results in the existence of ZES at that position\cite{kane}. This unpaired site no longer persists at the interface for $\theta=\pi/2$ as there is no modulation parameter $\Delta$ in hopping strength (t) at the interface bonds, which in turn gives rise to the disappearance of DW (of course ZESs residing in it) at the interface (though we are considering DW configuration). For $\theta=\pi/3$, however, one unpaired site at the interface in the strong coupling limit $|\Delta/t|=2$ also exists. Unlike $\theta=\pi,~\pi/3$, although there appears $\Delta$ term at the interface bonds, one can still notice the absence of zero modes for $\theta=\pi/4$ at the maximally dimerized limit $|\Delta/t|=\sqrt{2}$ as can be noticed from the low-energy spectra of Fig.\ref{numerical}(d). This analysis gives the physical intuition of having ZES at the interface for $\theta=\pi,~\pi/3$ and not observing ZES at the DW position for $\theta=\pi/2,~\pi/4$.

To support the above numerical analysis, we now focus on studying the same analytically.

\subsection{Analytical Study Considering Effective Dirac Hamiltonian: Jackiw-Rebbi Method}
In the seminal work by JR\cite{rebbi}, one-dimensional Dirac fermions are coupled with a solitonic field. The solitonic field was introduced in the low-energy effective Dirac Hamiltonian as a position-dependent mass term, the specific structure of which determines the topology of the system. A change of sign in the mass term about the DW yields localized ZES at its position. The topological defect of the solitonic field is known as kink. Polyacetylene is the condensed-matter realization of such
a situation at any domain wall junction between two possible dimerizations ($\Delta/t<0$ and $\Delta/t>0$)\cite{su1,su2}. The general recipe of the JR method is as follows: consider a Dirac equation, make the mass position-dependent (spatially inhomogeneous), then change its sign by introducing a topological defect and the result in the localization of trapped ZES on the defect (for details see Ref.\cite{jackiw}). Using this mechanism, we aim to investigate analytically the properties of ZESs in the presence of DW for different commensurate frequencies.

\subsubsection{ Case-I:~$\theta=\pi$}
This case resembles the usual SSH model which is two atom basis unit cells\cite{kar,mandal}. The corresponding single-particle Bloch Hamiltonian in momentum space takes the following form
\begin{equation}\label{5}
\mathcal{H}(k)=
\begin{pmatrix}
0 & P_{k} \\
 P_{-k} & 0
\end{pmatrix},
\end{equation}
where, $P_{k}=(t+\Delta)+ (t-\Delta)e^{-ik}$. It can be rewritten using the notion of Pauli matrices $\sigma_{i}$ as
\begin{equation}\label{5a}
\mathcal{H}(k)=[(t+\Delta)+(t-\Delta)cos(k)]\sigma_{x}+(t-\Delta)sin(k)\sigma_{y}
\end{equation}
and dispersion relation reads
\begin{equation}\label{6}
E(k)_{\pm}=\pm\sqrt{2}\sqrt{t^2+\Delta^2+(t^2-\Delta^2)cosk}
\end{equation} 
which exhibits a band gap at the edge of the first Brillouin zone (FBZ) at $k=\pm\pi/a$ for $\Delta/t\ne 0$ and has a magnitude 2$|\Delta|$\cite{comment} while gap closing at this FBZ boundary occurs for $\Delta/t=0$. Apart from this, the system is also gapless at $k=0$ for $t=0$. In addition to gap closing, here $\Delta/t=0$ is also a topological phase transition point (TPT)\cite{mandal,kar}. One can notice Dirac-like dispersion near this point\cite{bernevig}. To facilitate the understanding of low-energy physics for our model, close to the gap closing and near the phase transition point, the effective Hamiltonian reads
\begin{equation}\label{7}
\mathcal{H}(\hat{p})=-\frac{a(t-\Delta)}{\hbar}\hat{p}\sigma_{y}+M_{0}\sigma_{x}\\+\mathcal{O}(\hat{p}^2)
\end{equation}
where, $k\rightarrow -\pi/a+\hat{p}/\hbar$ (with $\hat{p}<<1$)\cite{comment1}, in which $\hat{p}$ and $\hbar$ denoting the momentum operator and the reduced Planck constant respectively and mass term $M_{0}~(=(t+\Delta)-(t-\Delta)=2|\Delta|$) describes the energy gap. As momentum $\hat{p}$ represents the derivative along the coordinate $x$ of the lattice (i.e., $\hat{p}=-i\partial_{x}$), thus one can have a continuum limit and the above Eq.(\ref{7}) taking the form of 
\begin{equation}\label{8}
\mathcal{H}(x)=iv\partial_{x}\sigma_{y}+M(x)\sigma_{x}\\+\mathcal{O}(\hat{p}^2)
\end{equation}
where $v=\frac{a(t-\Delta)}{\hbar}$ denotes a velocity. Notice that, the change in hopping parameters along the lattice, in the continuum limit, enforces the mass term to become spatially dependent $M(x)\propto\Delta(x)$. Therefore, the mass profile depends on a spatially dependent staggered hopping $\Delta(x)$, the typical choice of which is defined as $\Delta(x)=\Delta_{0}\tanh\Big[\frac{x-x_{0}}{\xi/a}\Big]$ centered at site $x=x_{0}$. Now, $\Delta(x)$ should satisfy $\lim \limits_{x-x_{0}\to-\infty}\Delta(x)=-\Delta_{0}$ and  $\lim \limits_{x-x_{0}\to\infty}\Delta(x)=\Delta_{0}$. For $\Delta_{0}>0$, two constant fields $\pm\Delta_{0}$ represent the two phases for our chain (phase $A$ and phase $B$) and each texture holds a single localized state with energy $E = 0$\cite{dw,su1,su2}. The kink-shaped mass profile or soliton field $\Delta(x)$ interpolating between $\Delta_{0}$ and $-\Delta_{0}$ and creates a DW at $x=x_{0}$.
\begin{figure}
     \vskip -.4 in
   \begin{picture}(100,100)
     \put(-90,0){
  \includegraphics[width=.54\linewidth, height=1.15 in]{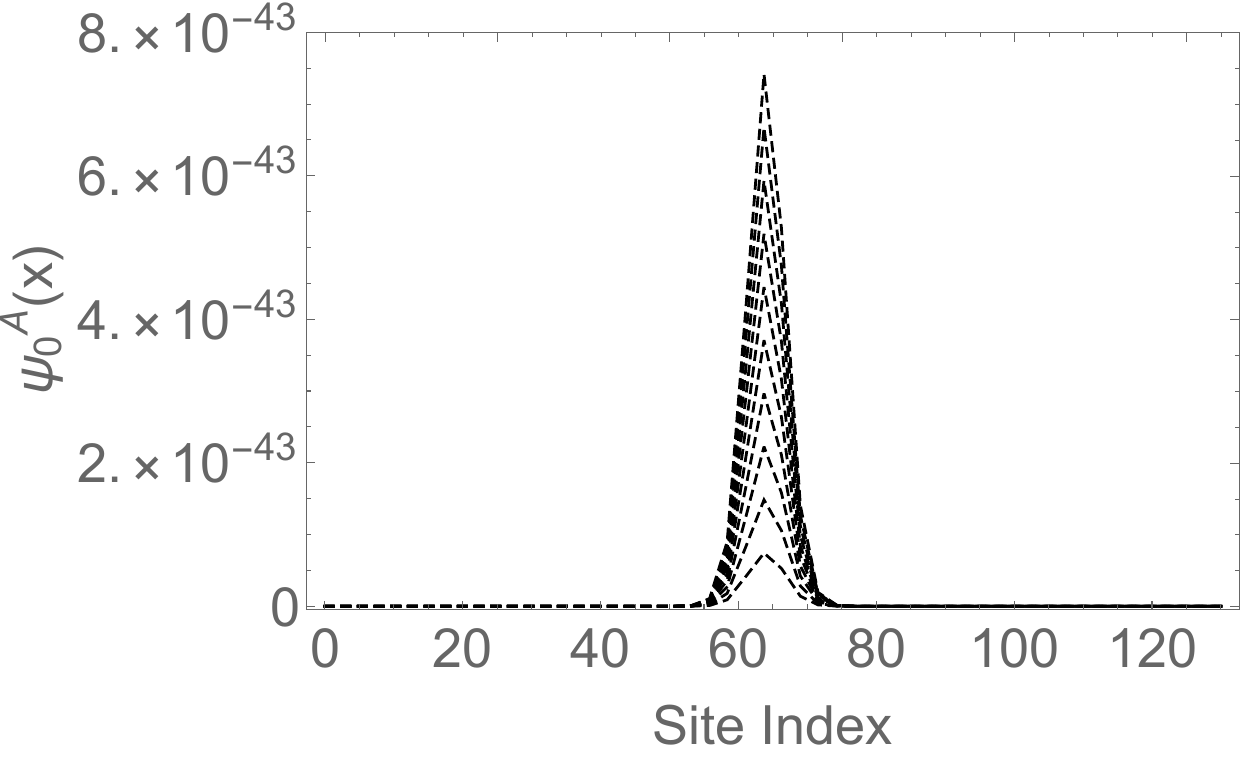}
  \includegraphics[width=.52\linewidth, height=1.15 in]{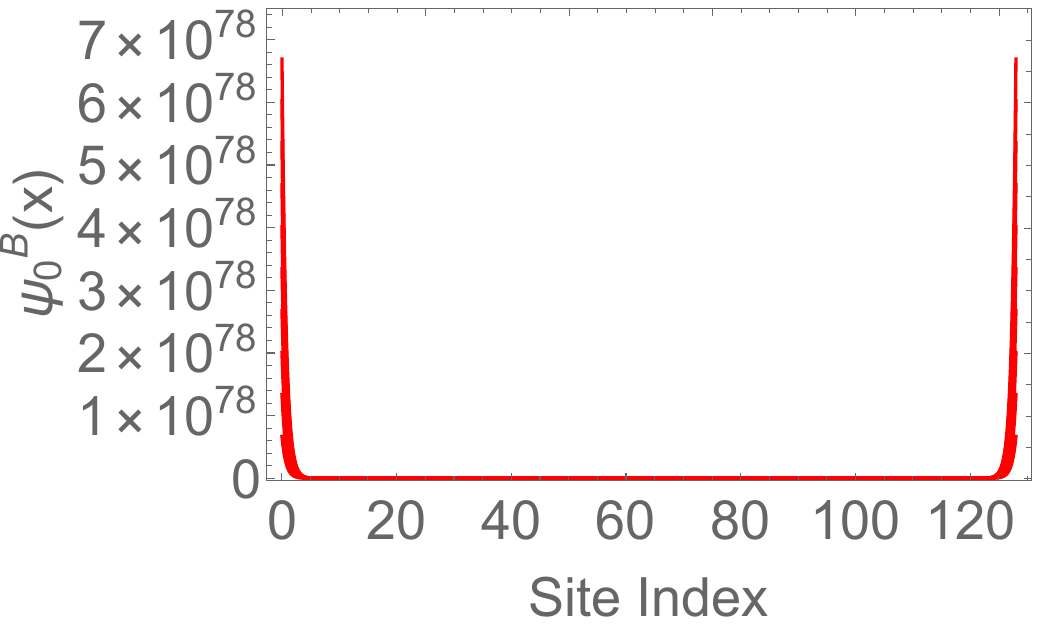}}
     \put(-40,60){(a)}
     \put(100,60){(b)}
    \end{picture}
\caption{Analytical spectrum of zero modes for generalized SSH chain with $\theta=\pi$ for $\psi_{0}^{A}(x)$ (a) and $\psi_{0}^{B}(x)$ (b).  The one $\psi_{0}^{A}(x)$ is localized at the DW position and the other $\psi_{0}^{B}(x)$ is hybridized ZES which resides at the two ends of the chain. In the plot, we considered ten different values of the normalization constant $N$ and $N^{\prime}$. We get peaks with less magnitudes for smaller values of normalization constant and the peak value in (b) exceeds unity due to an unnormalized solution. We set, $\Delta_{0}=0.7$, $L=128$ and $v=0.5$.} 
\label{fig4}
\end{figure}

Eq.(\ref{8}) gives two zero-energy solutions for the eigen equation $\mathcal{H}(x)\psi(x)=E\psi(x)$ and we define the two components of wave function $\psi(x)$ corresponding to these solutions as $\psi_{0}^{A}(x)$ and $\psi_{0}^{B}(x)$. The equation for these eigenfunction with energy $E_{0}=0$ is given by
\begin{equation}\label{9}
\mathcal{H}(x)\psi_{0}^{A}(x)=[M(x)+v\partial_{x}]\psi_{0}^{A}(x)=0
\end{equation}
\begin{equation}\label{9a}
\mathcal{H}(x)\psi_{0}^{B}(x)=[M(x)-v\partial_{x}]\psi_{0}^{B}(x)=0
\end{equation}
which gives the (unnormalized) solution as
\begin{equation}\label{10}
\psi_{0}^{A}(x)=N\cosh\Big(\frac{a(x_{0}-x)}{\xi}\Big)^{-\frac{\Delta_{0}\xi}{av}}
\begin{bmatrix} 1\\0 \end{bmatrix}
\end{equation}
\begin{equation}\label{10a}
\psi_{0}^{B}(x)=N^{\prime}\cosh\Big(\frac{a(x_{0}-x)}{\xi}\Big)^{\frac{\Delta_{0}\xi}{av}}
\begin{bmatrix} 0\\1 \end{bmatrix}
\end{equation}
with $N$ and $N^{\prime}$ refers to the normalization constants. Notice that, in Eq.(\ref{10}), zero mode $\psi_{0}^{A}(x)$ is reside at the DW position and decays exponentially as $x-x_{0}\rightarrow\pm\infty$. However, Eq.(\ref{10a}) shows that the zero mode, $\psi_{0}^{B}(x)$ localized at the single boundary. The graphical representation for these two solutions is depicted in Fig.\ref{fig4}. The two zero modes, $\psi_{0}^{A}(x)$ and $\psi_{0}^{B}(x)$ give the probability amplitudes of finding the electron on $A$ and $B$ sublattices (carbon atoms for polyacetylene), respectively. One zero mode ($\psi_{0}^{A}(x)$) is bound at the DW position which corresponds to a zero-energy soliton/kink mode\cite{yang} and is very well localized near the $A$-sublattices around $x=x_{0}$ while the other ($\psi_{0}^{B}(x)$) is at one end of the chain (maybe at $x=0$ or $L$) and on the $B$-sublattics. Therefore, zero modes are restricted to a single sublattice and they do have nonzero amplitude in more than a single but alternative sites when hopping periodicity is enhanced as we studied later. Interestingly, as soliton comes from a variation of the dimerization $M(x)$, it represents a chiral mode and is topologically robust\cite{yang}.
\subsubsection{Case-II:~$\theta=\pi/2$}
The lattice periodicity increases twice here for $\theta=\pi/2$ with the number of sublattices becoming four. The Bloch Hamiltonian in the $k$-space for the periodic chain reads
\begin{equation}\label{11}
\mathcal{H}(k) = 
\begin{pmatrix}
0 & (t+\Delta) & 0 & te^{-4ik} \\
(t+\Delta) & 0 & t & 0 \\
0 & t &0 & (t-\Delta) \\
te^{4ik} & 0 & (t-\Delta) & 0
\end{pmatrix}.
\end{equation}
In terms of Pauli matrices, Eq.(\ref{11}) takes the following form:
%\small
\begin{multline}
 \mathcal{H}(k)=t\sigma_{0}\otimes\sigma_{x}+\Delta\sigma_{z}\otimes\sigma_{x}+\frac{t}{2}(1+\cos4k)\nonumber\sigma_{x}\otimes\sigma_{x}\\+ \frac{t}{2}(1-\cos4k)\sigma_{y}\otimes\sigma_{y}+\frac{t}{2}\sin4k(\sigma_{x}\otimes\sigma_{y}+\sigma_{y}\otimes\sigma_{x})
 \end{multline}
%\end{align}}

with dispersion can be obtained as
%\begin{widetext}
\begin{align}\label{12}
E(k)=\pm\sqrt{2t^2+\Delta^2\pm t\sqrt{2t^2+6\Delta^2+2(t^2-\Delta^2) \cos4k}}
\end{align}
which indicates the gap closing point at the boundary of FBZ at  $k=\pm\pi/4a$ for $|\Delta/t|=\sqrt{2}$ or at $k=0$ for $|\Delta/t|=0$. However, the band gap at the boundary of FBZ becomes  $M_{0}=\pm(\sqrt{2}t\pm\Delta)$. Since $|\Delta/t|=\sqrt{2}$ represents TPT here for $\theta=\pi/2$, we can perform low-energy description for this cases around $k=\pm\pi/4a$ and arrive at
\begin{multline}\label{low}
 \mathcal{H}(\hat{p})=t(\sigma_{0}\otimes\sigma_{x}+\sigma_{y}\otimes\sigma_{y})+\Delta\sigma_{z}\otimes\sigma_{x}+\\\frac{-2at\hat{p}}{\hbar}(\sigma_{x}\otimes\sigma_{y}+\sigma_{y}\otimes\sigma_{x})+\mathcal{O}(\hat{p}^2)
 \end{multline}
where, $k=-\pi/4a+\hat{p}/\hbar$. In the continuum limit, the above Eq.(\ref{low}) becomes
\begin{multline}\label{continuum}
 \mathcal{H}(\hat{x})=t(\sigma_{0}\otimes\sigma_{x}+\sigma_{y}\otimes\sigma_{y})+\Delta\sigma_{z}\otimes\sigma_{x}+\\\frac{i2at}{\hbar}\partial_{x}(\sigma_{x}\otimes\sigma_{y}+\sigma_{y}\otimes\sigma_{x})+\mathcal{O}(\hat{p}^2)
 \end{multline}

\begin{figure}
  \vskip -.4 in
   \begin{picture}(100,100)
     \put(-90,0){
  \includegraphics[width=.54\linewidth, height=1.15 in]{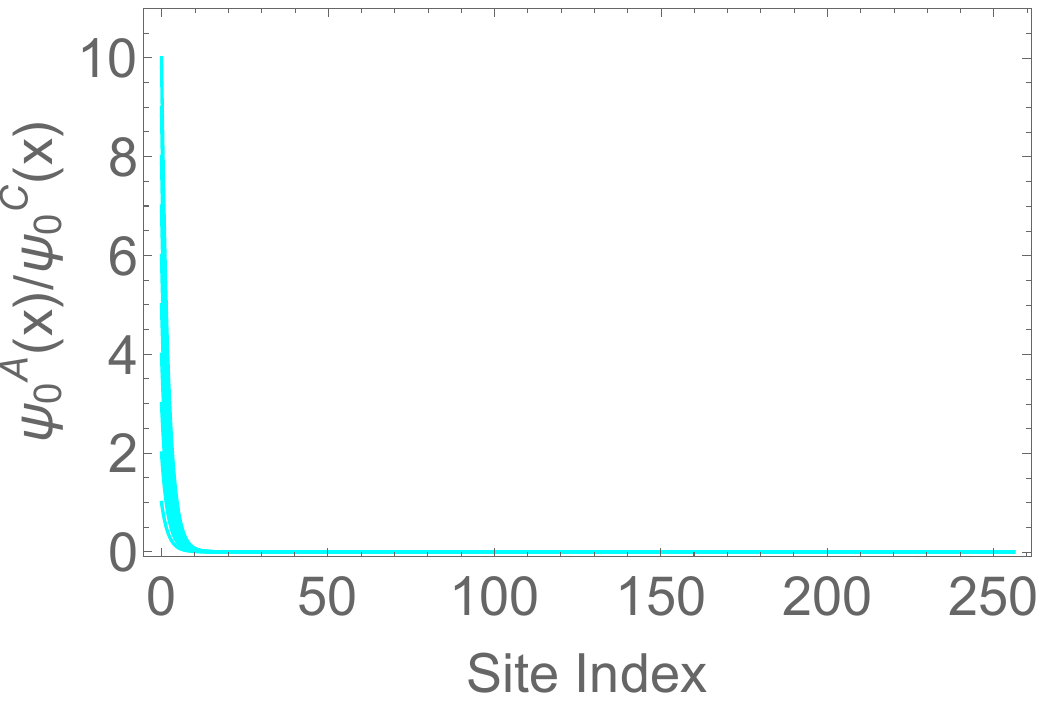}
  \includegraphics[width=.52\linewidth, height=1.15 in]{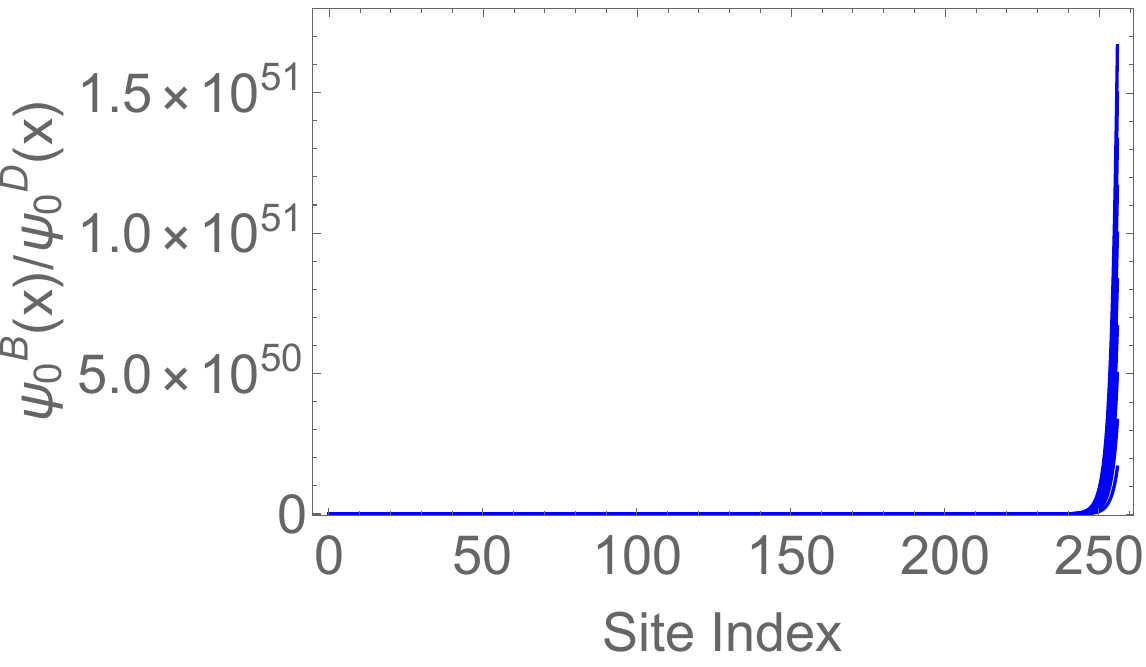}}
     \put(-40,60){(a)}
     \put(100,60){(b)}
    \end{picture}
\caption{Analytical spectrum of ZESs for generalized SSH chain with $\theta=\pi/2$ for (a) $\psi_{0}^{A}(x)$ or $\psi_{0}^{C}(x)$ and (b) $\psi_{0}^{B}(x)$ or $\psi_{0}^{D}(x)$(b). Two ZESs are bound to the two ends of the chain, and no DW state is found. Here other choosing parameters are $\Delta_{0}=0.7$, $L=256$ and $v=0.5$.}
\label{fig5}
\end{figure}

Obeying eigen equation $H(x)\psi(x)=0$ with  $\psi(x)=\Big(\psi_{0}^{A}(x)$, $\psi_{0}^{B}(x)$, $\psi_{0}^{C}(x)$, $\psi_{0}^{D}(x)\Big)^T$ and each components in $\psi(x)$ provide the probability density on the $A$, $B$, $C$ and $D$ sublattices respectively. It is convenient to obtain the two solutions, namely $\psi_{0}^{A}(x)$ and $\psi_{0}^{C}(x)$, by setting $\psi_{0}^{B}(x)$=$\psi_{0}^{D}(x)$=0. The solution for these two component eigenfunctions read as:
\begin{widetext}
\begin{equation}\label{13}
\psi_{0}^{A}(x)=\psi_{0}^{C}(x)=cExp\Bigg[-\frac{\Delta_{0}^2\Big(\frac{-\xi Arc\tanh(\tanh(\frac{a(x_{0}-x)}{\xi}))}{a}+\frac{\xi \tanh(\frac{a(x_{0}-x)}{\xi})}{a}\Big)}{v}\Bigg]
\begin{bmatrix} 1\\0 \end{bmatrix}
\end{equation}
%\end{widetext}
While the setting of $\psi_{0}^{A}(x)$=$\psi_{0}^{C}(x)$=0 gives the solution for other two components as
%\begin{widetext}
\begin{equation}\label{14}
\psi_{0}^{B}(x)=\psi_{0}^{D}(x)=c^{\prime} Exp\Bigg[\frac{\Delta_{0}^2\Big(\frac{-\xi Arc\tanh(\tanh(\frac{a(x_{0}-x)}{\xi}))}{a}+\frac{\xi \tanh(\frac{a(x_{0}-x)}{\xi})}{a}\Big)}{v}\Bigg]
\begin{bmatrix} 0\\1 \end{bmatrix}
\end{equation} 
\end{widetext}
with $c$ and $c^{\prime}$ denoting the normalization factors and $v=\frac{4at^2}{\hbar}$ is a velocity here for $\theta=\pi/2$. It is noticeable from Eq.(\ref{13}) that, the amplitude of the first and third components of the wave function is completely localized on $A$ and $C$ sublattices respectively, and decays exponentially away from the edges (left) while the amplitude of the other two components vanishes on $B$ and $D$ sublattices. In contrast, the ZES, while vanishing first and third components, for the second and fourth components have support on $B$ and $D$ sublattices respectively, and at the right end of the chain (notice Eq.(\ref{14})). This indicates that ZESs are restricted to two but nonconsecutive sublattices. In other words, one ZES is bound to $A$ and $C$ sites while the other at rest two alternate sites. To support this, we make a graphical illustration based on Eqs.(\ref{13}) and (\ref{14}) in Fig.\ref{fig5}. The plot shows that $A$ and $C$ components of the wave function are localized at the left end of the chain and the chain shows a peak at the right end for $B$ and $D$ components.

As we consider the case of $\theta=\pi/3$, one ZES trapped at the topological DW and the other at the one end of the chain. Moreover, the two ZESs for $\theta=\pi/4$ are shown to be localized at the two ends of the chain. In order to maintain the main flow of the paper to the reader, we keep the corresponding analytical study in {\bf Appendix~A}. The above analytical prediction and its comparison with the numerical analysis given in Fig.\ref{zes}, exhibit an excellent agreement.

In general, the interface between the two dimerized phases of periodically modulated SSH chain results in the emergence of ZES at DW position only for $\theta=\frac{\pi}{2s+1}$ for zero or an integer value of $s$ whereas, no states doesn't show localization at DW for $\theta=\frac{\pi}{2s}$ with nonzero and integer $s$ value. The ZES residing on DW is sometimes known as DW solitons having fractional charge $\pm e/2$. Such typical ZES were first proposed by JR\cite{rebbi} and are called JR modes. Our model with $\theta=\frac{\pi}{2s+1}$ gives only a physical realization of the JR modes. We now add here that, the edge of the SSH chain holds a dangling bond as the cut occurs at the strong bond of the chain\cite{dw}. This mid-gap state sites are visible at zero energy and can be empty/occupied by an electron (spinless here). The mid-gap state, in the absence of an electron, has an extra charge $+e/2$ (that of the isolated cation at the edge) while it has a charge $-e/2$ (that of the cation+electron=$+e/2-e=+e/2$) for occupied cases. This is a lucid mechanism of charge fractionalization discovered by JR\cite{rebbi,dw}.

We should mention here a related Shiozaki-Sato-Gomi model which considers constant NN hopping but staggered on-site potentials and respects a nonsymmorphic chiral symmetry. This model, in contrast to the SSH model, features unpaired ZESs at the position of a smooth DW and no end states reside at sharp interfaces with vacuum because of breaking nonsymmorphic chiral symmetry at the boundary. Interestingly, DW solitons for this model can carry irrational charge $e=\frac{1}{2}(1\pm f)$, with $f$ denoting the breaking of the chiral symmetry\cite{ssg}.

\section{Localization Property of Zero Energy Edge and DW states}\label{sec5}
Here we measure the amount of localization of the edge, and DW states in our generalized SSH system. The inverse participation ratio (IPR)\cite{scollon,roy} can estimate that information.  The IPR for the $m$-th eigenstate, $\phi_{m}^{i}$, can be written as $$\mathrm{IPR_{m}}=\sum_{i=1}^{L}|\phi_{m}^{i}|^4,$$ which gives 1 (0) for a maximally localized (extended) states for the large $L$ limit\cite{roy}.

\begin{figure}
  \vskip -.4 in
   \begin{picture}(100,100)
     \put(-90,0){
  \includegraphics[width=.54\linewidth, height=1.15 in]{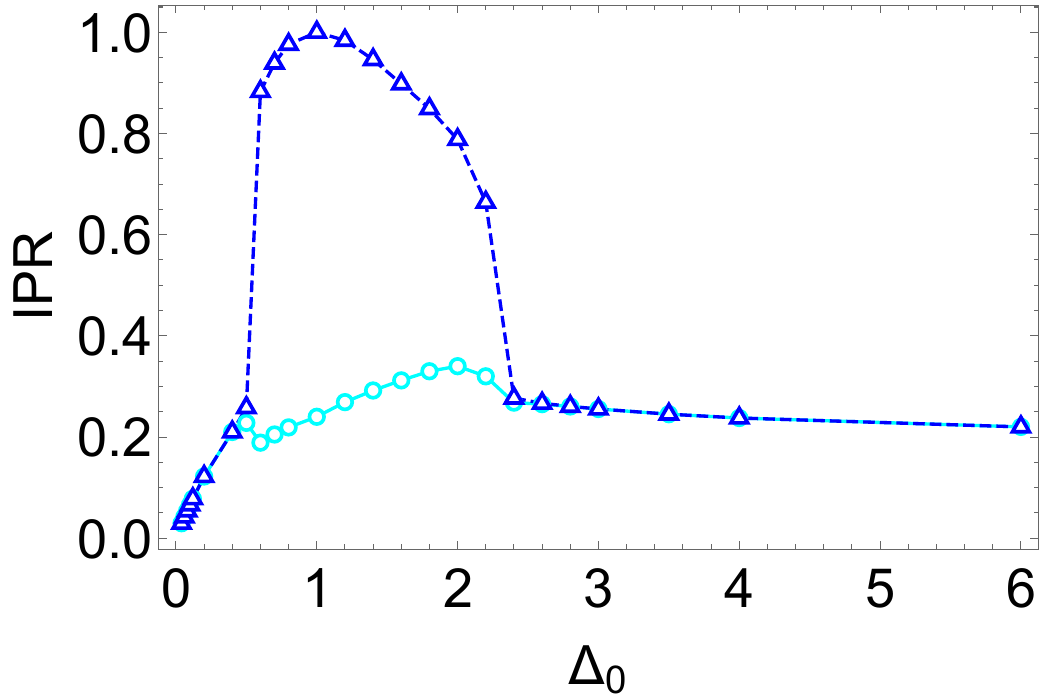}
  \includegraphics[width=.52\linewidth, height=1.15 in]{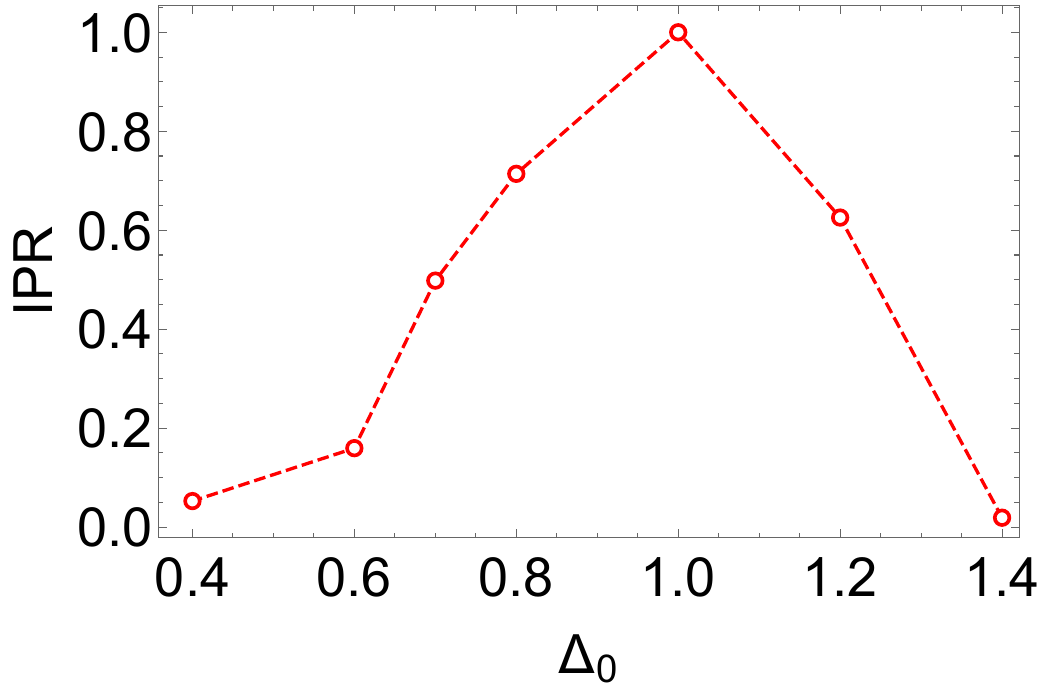}}
     \put(-15,60){(a)}
     \put(85,60){(b)}
     \end{picture}\\
     \begin{picture}(100,100)
     \put(-90,0){
      \includegraphics[width=.54\linewidth, height=1.15 in]{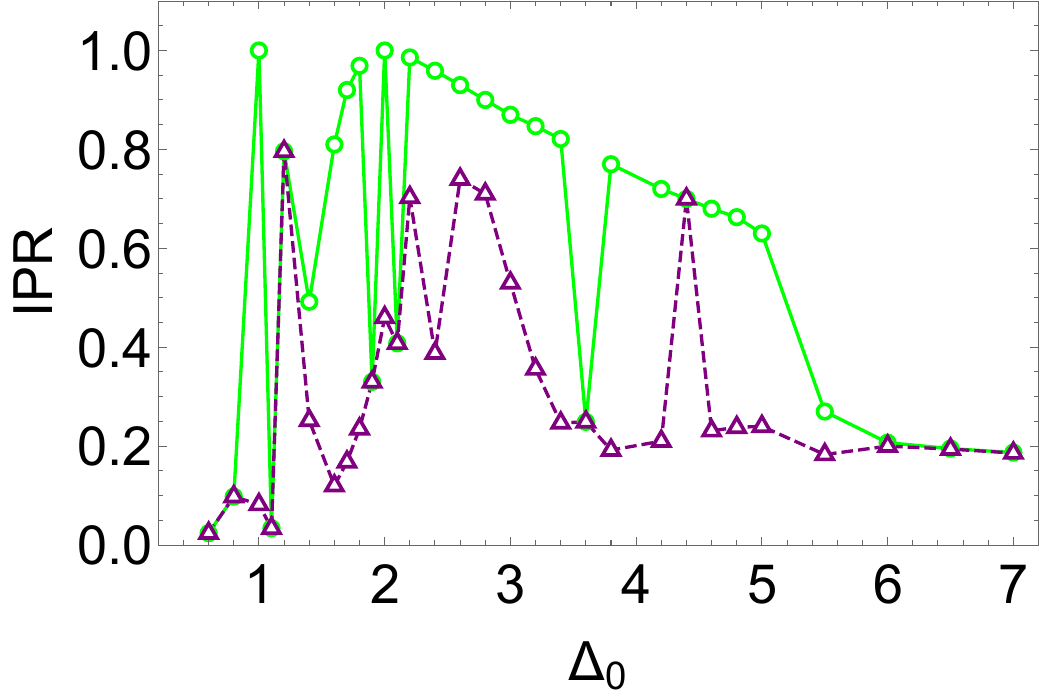}
  \includegraphics[width=.52\linewidth, height=1.15 in]{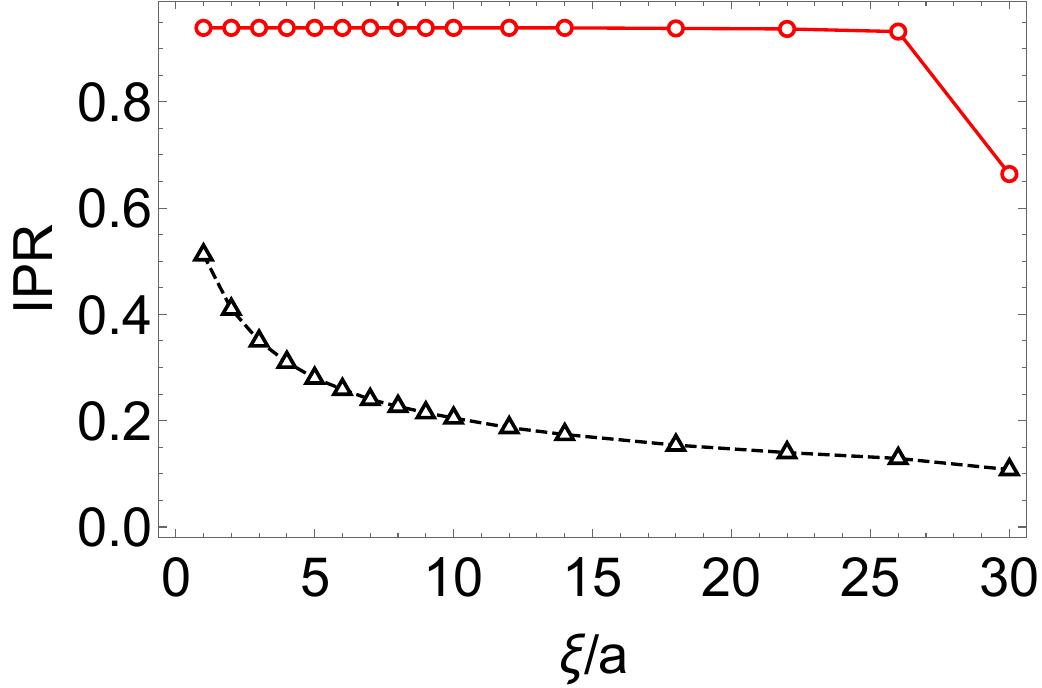}}
     \put(20,65){(c)}
     \put(100,60){(d)}
    \end{picture}
\caption{IPRs for mid-gap ZESs in the presence of DW for $\theta$~(a) $\pi$ with $L=128$, (b) $\pi/2$ with $L=256$, (c) $\pi/3$ with $L=360$. In these three cases, we fix $\xi/a=10$. In (d), we plot IPR with $\xi/a$ for $\theta=\pi$ when $\Delta_{0}=0.7$ and $L=128$. The blue and cyan lines in (a), the green and purple lines in (c), and the red and black lines in (d) are for zero energy edge and DW states respectively. }
\label{ipr}
\end{figure}

The localization of ZESs away from the edges scaled as $\sim|\frac{t-\Delta_{0}}{t+\Delta_{0}}|$ which amounts in a lesser localization for smaller $\Delta_{0}(\sim 0.5)$ as well as larger $\Delta_{0}(>2.4)$. This rather, however, indicates that up to $\Delta_{0}(\sim 0.5)$ two degenerate ZESs are hybridized states having symmetric and anti-symmetric pairs which have a localization both on edge and DW position and beyond that limit of $\Delta_{0}$ ZESs become perfectly localized. Among the two, one exhibits increasing IPRs near the one end of the chain while a slower growth in IPRs occurs near the location of DW. Both the edge and DW states have a maximum localization at a strong coupling limit $|\Delta_{0}/t|=1$. Changing $\Delta_{0}$ from strong coupling limit indicates the advent of extended states and gives rise to hybridized ZESs (notice Fig.\ref{ipr}(a)). As there is no state bound at DW for $\theta=\pi/2$, IPRs of ZESs tend to that found in the absence of DW\cite{mandal1} (Fig.\ref{ipr}(b)). However, for $\theta=\pi/3$ in addition to the smaller and larger $\Delta_{0}$, ZESs become hybridized states for some intermediate $\Delta_{0}$ values (notice Fig.\ref{ipr}(c)). The ZES traped on the DW shows maximum localization for sharp DW however, it moves towards the extended states as the width of DW increases. Typically, the topological DW states display extended nature as $\sim1/\xi$. On the other hand, the zero-energy edge state exhibits extended property only for large $\xi$ limit (see Fig.\ref{ipr}(d)).

\section{Summary and conclusions}\label{sec6}
In this paper, we have tried to study the behavior of ZESs of the SSH model in the presence of DW configuration, concurrently, adding intuition into the same while addressing additional periodic modulation in the hopping parameter. The ZESs show noteworthy evolutions with the variation of hopping periodicity. It is shown both numerically and analytically that in the presence of a static DW that has been placed at an interface between two dimerized phases of a finite chain, one ZES gets depleted which further appears at the position of the DW as soon as $\theta=\frac{\pi}{2s+1}$ for zero or an integer value of $s$ while no ZES persists at DW for $\theta=\frac{\pi}{2s}$ with nonzero and an integer $s$ value. The ZESs bound about the mass DW are familiar as DW solitons carrying fractional charge $\pm e/2$. Such typical ZESs are called JR modes. Our model with $\theta=\frac{\pi}{2s+1}$ realizes the JR modes physically. These fractional JR modes do have an exotic phase because of their excitations and can be a possible candidate for implementing topological quantum computation\cite{nayak}.

Unlike the SSH chain, here in our model ZESs have support not only in a single sublattice rather they do have support in more than a single but also in alternative sublattices (their restriction on sublattices entirely depends on the hopping periodicity. As hopping periodicity increases, their restriction on sublattices increases). In general, for $\theta=\frac{\pi}{2s+1}$ ZESs are bounded in an odd number of sublattices whereas, they do have support in an even number of sites for $\theta=\frac{\pi}{2s}$ and site alteration for getting zero or nonzero amplitude of ZESs is noticeable for $s\ge 1$ (or $\theta<\pi$). We studied the IPRs to analyze the localization of edge and DW states staying at zero energy. For all the commensurate frequencies considered herein, the localization is being suppressed for both the smaller and larger DW amplitude $\Delta_{0}$. The maximally localized topological DW state is obtained at sharp DW that loses its strength gradually as DW becomes smooth. Typically, localized DW state scaled to extended states as $\sim1/\xi$. Moreover, the edge state exhibits an extended nature when $\xi$ is considerably large.

The noteworthy evolution of ZESs with hopping periodicity for generalized SSH chain can be verified in cold atom setup within optical lattices\cite{a1}, or specially designed graphene nanoribbons\cite{a2} or maybe in topological acoustic systems\cite{a3}. In the future, we have a plan to study the out-of-equilibrium behavior of such hopping modulated SSH model, subjected to a quantum quench which leads to an effective metal-insulator transition for $\theta=\pi$\cite{a4}. It would also be interesting to study a Floquet analysis for a high-frequency periodic quench\cite{a5} and explore the competition between the time-periodic driving and the topology in our generalized SSH model.

\section*{Acknowledgements}
The author thanks S. Kar and S. Mandal (IOP, Bhubaneswar) for the fruitful discussions. The author acknowledges financial support from DST-SERB (ANRF), Government of India under grant no. CRG/2022/002781.

\section*{Appendix~A:~Case of $\theta=\pi/3$ and $\pi/4$}
The Bloch Hamiltonian for $\theta=\pi/3$ is given in Eq.(\ref{15}) in which $z=e^{-6ik}$\cite{mandal}. The unit cell now consists of six sublattices.
\begin{align}\label{15}
H(k) =
\begin{pmatrix}
0 & (t+\Delta) & 0 & 0 & 0 & (t+\frac{\Delta}{2})z \\
(t+\Delta) & 0 & (t+\frac{\Delta}{2}) & 0 & 0 & 0 \\
0 & (t+\frac{\Delta}{2}) & 0 & (t-\frac{\Delta}{2}) & 0 & 0 \\
0 & 0 & (t-\frac{\Delta}{2}) & 0 & (t-\Delta) & 0 \\
0 & 0 & 0 & (t-\Delta) & 0 & (t-\frac{\Delta}{2}) \\
(t+\frac{\Delta}{2})z^* & 0 & 0 & 0 & (t-\frac{\Delta}{2}) & 0 \\
\end{pmatrix}
\tag{A1}\end{align}

\begin{figure}
   \vskip -.4 in
   \begin{picture}(100,100)
     \put(-90,0){
  \includegraphics[width=.54\linewidth, height=1.15 in]{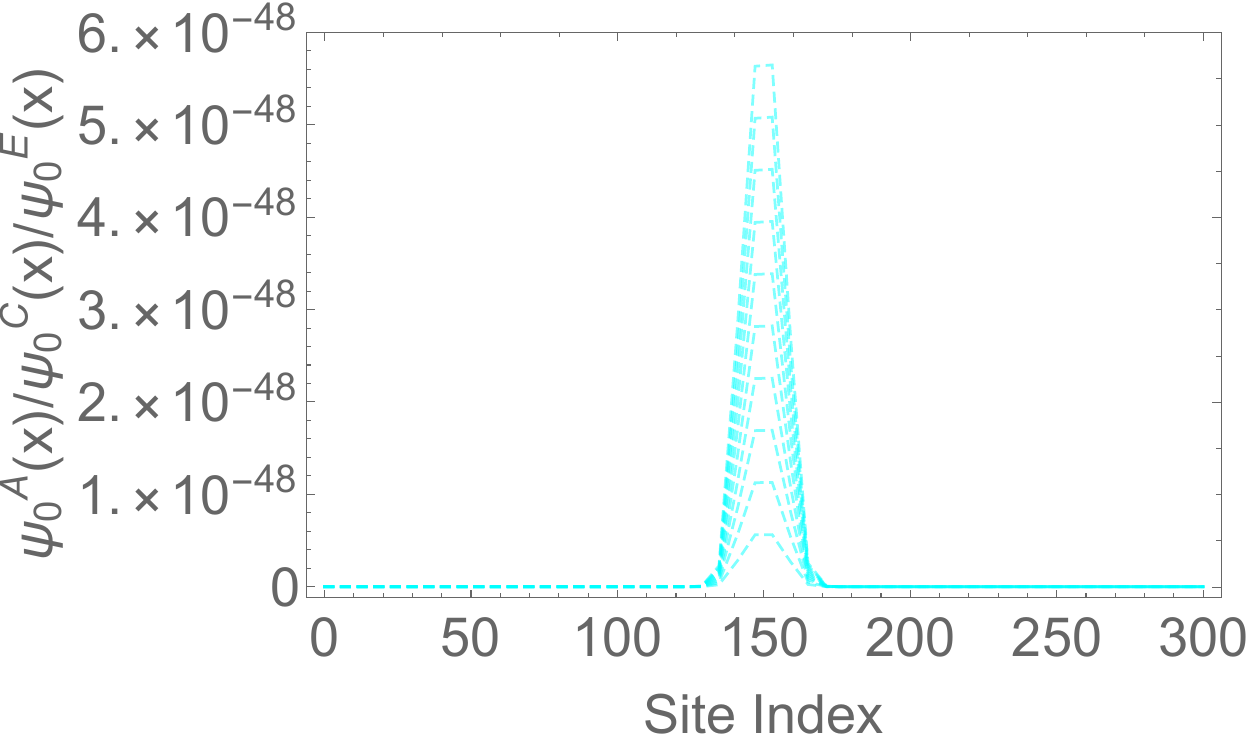}
  \includegraphics[width=.52\linewidth, height=1.15 in]{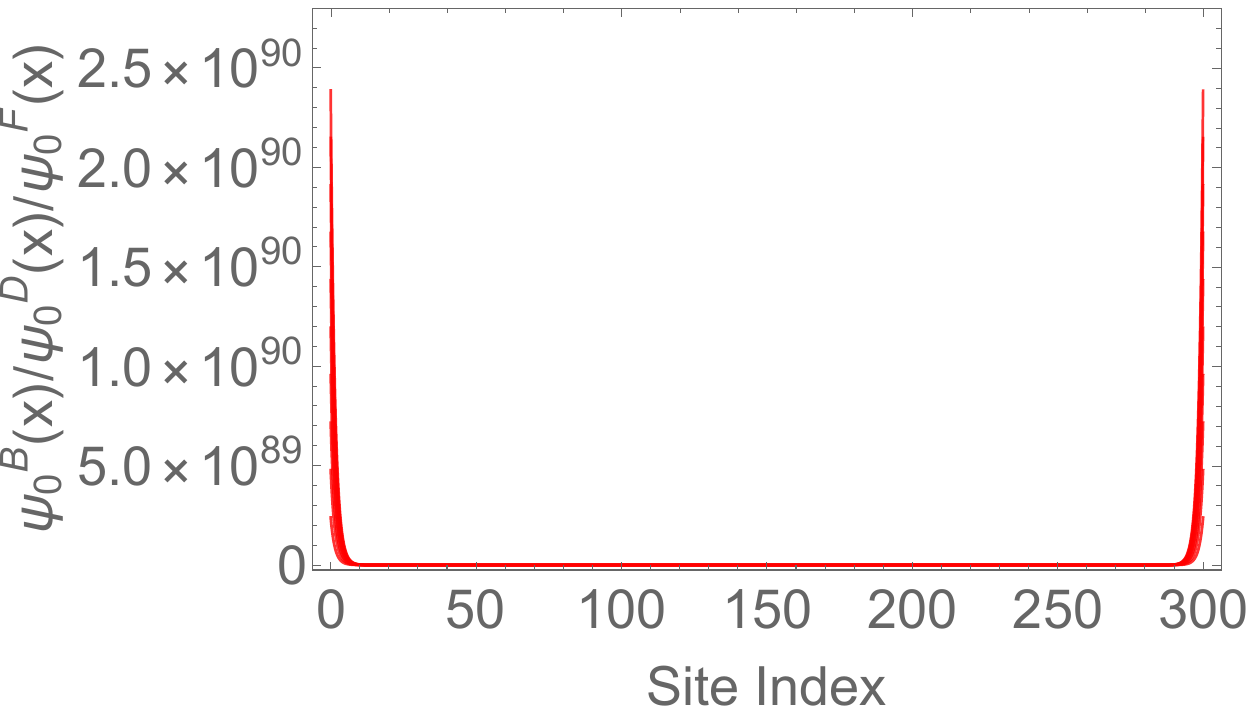}}
     \put(-40,60){(a)}
     \put(100,60){(b)}
    \end{picture} 
\caption{Analytical spectrum of zero modes for $\theta=\pi/3$ for (a) $\psi_{0}^{A}(x)$ or $\psi_{0}^{C}(x)$ or $\psi_{0}^{E}(x)$ and (b) $\psi_{0}^{B}(x)$ or $\psi_{0}^{D}(x)$ or $\psi_{0}^{F}(x)$. Here we set $\Delta_{0}=0.7$, $L=300$ and $v=0.5$.} 
\label{piby3}
\end{figure}

Here one can notice the gap closing points at $|\Delta/t|=2/\sqrt{3}$ for $k=0$ while it occurs at the boundary of FBZ at $\Delta=0$ for $k=\pm\pi/6a$. At the boundary of FBZ, the band gap becomes $M_{0}=0$. Although, $|\Delta/t|=0,2/\sqrt{3}$ accounts for TPT here, we are limited in considering TPT at $k=\pm\pi/6a$ for the low-energy description. To do so, by considering $k=-\pi/6a+\hat{p}/\hbar$, the Eq.(\ref{15}) in continuum limiting cases becomes:
\begin{widetext}
\begin{align}\label{16}
H(x) =
\begin{pmatrix}
0 & (t+\Delta) & 0 & 0 & 0 & -(t+\frac{\Delta}{2})+\frac{6}{\hbar}(t+\Delta/2)\partial_{x} \\
(t+\Delta) & 0 & (t+\frac{\Delta}{2}) & 0 & 0 & 0 \\
0 & (t+\frac{\Delta}{2}) & 0 & (t-\frac{\Delta}{2}) & 0 & 0 \\
0 & 0 & (t-\frac{\Delta}{2}) & 0 & (t-\Delta) & 0 \\
0 & 0 & 0 & (t-\Delta) & 0 & (t-\frac{\Delta}{2}) \\
-(t+\frac{\Delta}{2})-\frac{6}{\hbar}(t+\Delta/2)\partial_{x} & 0 & 0 & 0 & (t-\frac{\Delta}{2}) & 0 \\
\end{pmatrix}
\tag{A2}\end{align}
\end{widetext}
with $\hat{p}=-i\partial_{x}$. The above Eq.(\ref{16}) provides six solutions. For, $\psi_{0}^{B}(x)$=$\psi_{0}^{D}(x)$=$\psi_{0}^{F}(x)=0$, the solutions for other three components of the wave functions take the form:
\begin{widetext}
\begin{equation}\label{17}
\psi_{0}^{A}(x)=\psi_{0}^{C}(x)=\psi_{0}^{E}(x)=c_{1}Exp\Bigg[-\frac{\Delta_{0}^3\xi\Big(2\log \cosh(\tanh(\frac{a(x_{0}-x)}{\xi}))-{\tanh^2(\frac{a(x_{0}-x)}{\xi})}\Big)}{2av}\Bigg]
\begin{bmatrix} 1\\0 \end{bmatrix}
\tag{A3}\end{equation}
%\end{widetext}
However, considering $\psi_{0}^{A}(x)$=$\psi_{0}^{C}(x)$=$\psi_{0}^{E}(x)=0$ leads to the other three solution as
%\begin{widetext}
\begin{equation}\label{18}
\psi_{0}^{B}(x)=\psi_{0}^{D}(x)=\psi_{0}^{F}(x)=c_{2} Exp\Bigg[\frac{\Delta_{0}^3\xi\Big(2\log \cosh(\tanh(\frac{a(x_{0}-x)}{\xi}))-{\tanh^2(\frac{a(x_{0}-x)}{\xi})}\Big)}{2av}\Bigg]
\begin{bmatrix} 0\\1 \end{bmatrix}
\tag{A4}\end{equation} 
\end{widetext}
with $c_{1}$, $c_{2}$ representing the normalization factors and $v=\frac{6}{\hbar}(t^3-\frac{3t\Delta^2}{4}-\frac{\Delta^3}{4})$ being the velocity. It is noticeable from Eqs.(\ref{17}) and (\ref{18}) that, the amplitudes of one ZES become nonzero in three alternative sites $A$, $C$, and $E$ while its strength vanishes at the rest three alternative sublattices and vice-versa. This suggests the limitation of ZESs to three sublattices such that three alternative sites see nonzero strength of them. The corresponding analytical plot is presented in Fig.\ref{piby3} and illustrates that the ZES which has support on $A$, $C$, and $E$ sublattices is bound at the DW position, however, we notice nonzero strength at the edges for ZES which has support on the other three alternative sites.

In a similar fashion, for $\theta=\pi/4$ the energy gap vanishes at FBZ boundary at $|\Delta/t|=\sqrt{2(2\pm\sqrt{2})}$ for $k=\pm\pi/8a$ or for $k=0$ at $|\Delta/t|=0$ and the energy gap $M_{0}=\pm\Big(\sqrt{2(2\pm\sqrt{2}})t\pm\Delta\Big)$ can be notice at FBZ boundary. Letting $k=-\pi/8a+\hat{p}/\hbar$, the effective Hamiltonian in the continuum limiting cases obtained as
\begin{widetext}
\begin{equation}\label{19}
H(x) =
\begin{pmatrix}
0 & (t+\Delta) & 0 & 0 & 0 & 0 & 0 & -(t+\frac{\Delta}{\sqrt{2}})+\frac{8}{\hbar}(t+\frac{\Delta}{\sqrt{2}})\partial_{x} \\
(t+\Delta) & 0 & (t+\frac{\Delta}{\sqrt{2}}) & 0 & 0 & 0 & 0 & 0 \\
0 & (t+\frac{\Delta}{\sqrt{2}}) & 0 & t & 0 & 0 & 0 & 0 \\
0 & 0 & t & 0 & (t-\frac{\Delta}{\sqrt{2}}) & 0 & 0 & 0 \\
0 & 0 & 0 & (t-\frac{\Delta}{\sqrt{2}}) & 0 & (t-\Delta) & 0 & 0\\
0 & 0 & 0 & 0 & (t-\Delta) & 0 & (t-\frac{\Delta}{\sqrt{2}} & 0 \\
0 & 0 & 0 & 0 & 0 & (t-\frac{\Delta}{\sqrt{2}}) & 0 & t \\
-(t+\frac{\Delta}{\sqrt{2}})-\frac{8}{\hbar}(t+\frac{\Delta}{\sqrt{2}})\partial_{x} & 0 & 0 & 0 & 0 & 0 & t & 0 \\
\end{pmatrix}
\tag{A5}\end{equation}
\end{widetext}

\begin{figure}
   \vskip -.4 in
   \begin{picture}(100,100)
     \put(-90,0){
  \includegraphics[width=.54\linewidth, height=1.15 in]{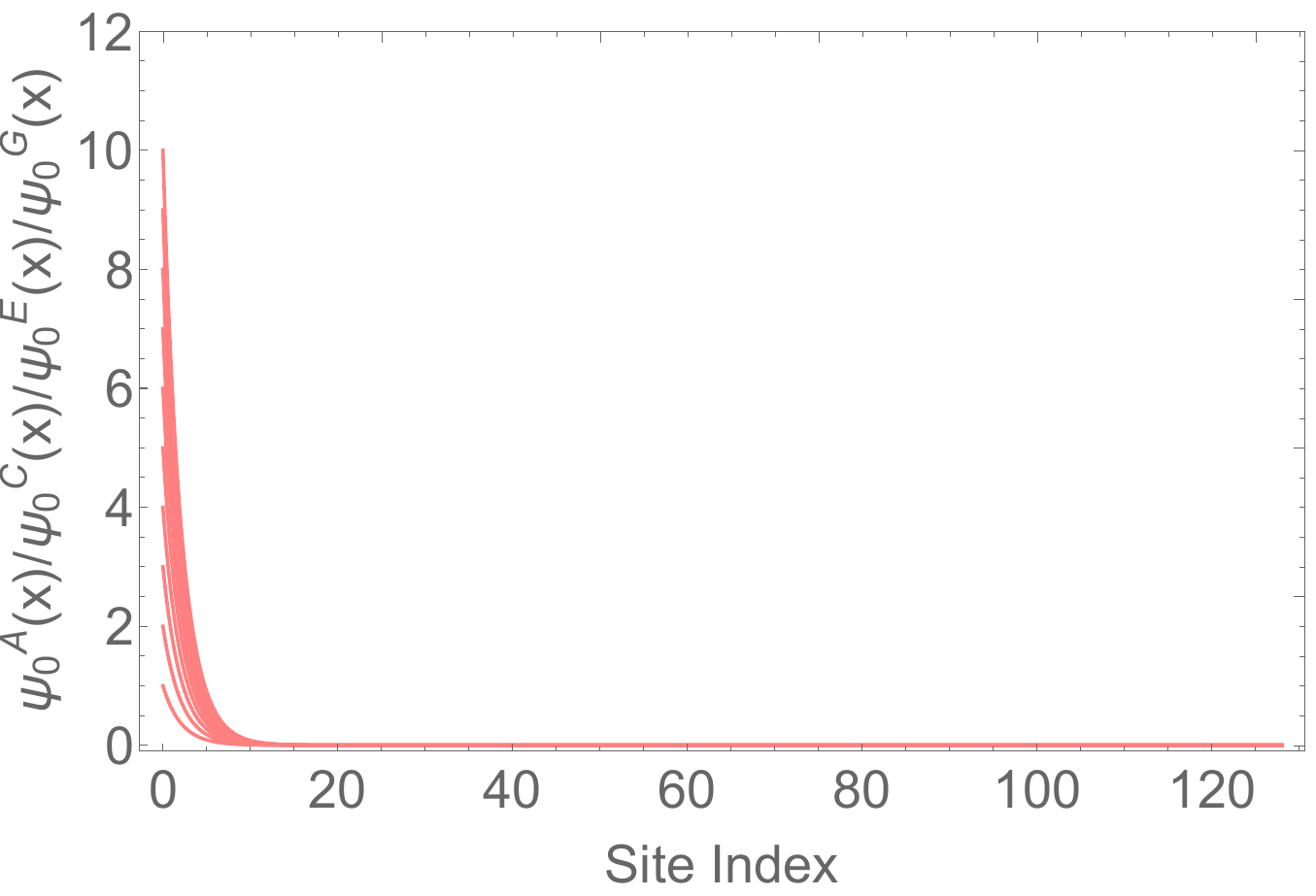}
  \includegraphics[width=.52\linewidth, height=1.15 in]{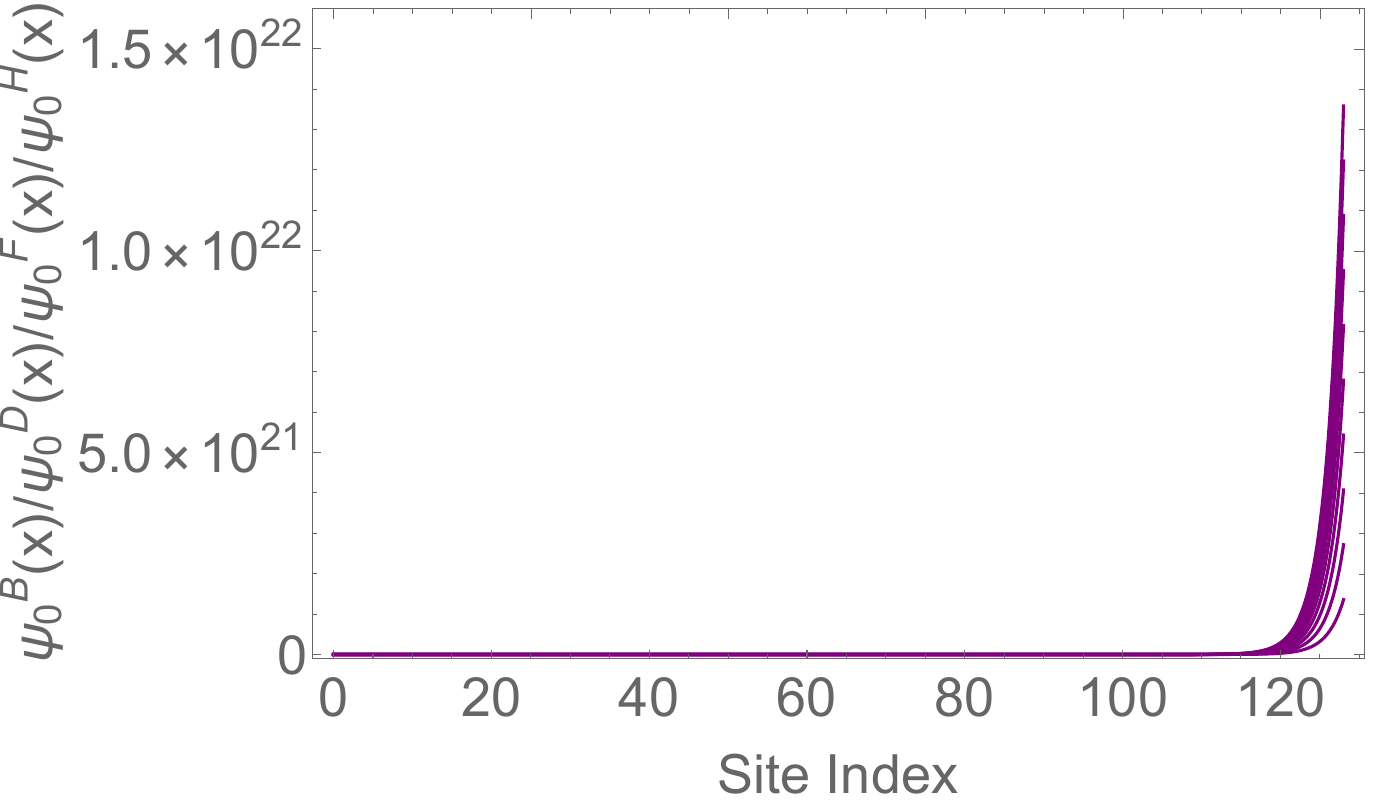}}
     \put(-40,60){(a)}
     \put(100,60){(b)}
    \end{picture} 
\caption{Analytical spectrum of zero modes for $\theta=\pi/4$ for (a) $\psi_{0}^{A}(x)$ or $\psi_{0}^{C}(x)$ or $\psi_{0}^{E}(x)$ or $\psi_{0}^{G}(x)$ and (b) $\psi_{0}^{B}(x)$ or $\psi_{0}^{D}(x)$ or $\psi_{0}^{F}(x)$ or $\psi_{0}^{H}(x)$. Here we set $\Delta_{0}=0.7$, $L=128$ and $v=0.5$.} 
\label{piby4}
\end{figure}

For eigenfunction solutions, we follow $H(x)\psi(x)=0$ with $\psi=\tiny{\Big(\psi_{0}^{A}(x),\psi_{0}^{B}(x),\psi_{0}^{C}(x),\psi_{0}^{D}(x),\psi_{0}^{E}(x),\psi_{0}^{F}(x),\psi_{0}^{G}(x),\psi_{0}^{H}(x)\Big)^T}$. One attains four solutions for $\psi_{0}^{A}(x),\psi_{0}^{C}(x),\psi_{0}^{E}(x),\psi_{0}^{G}(x)$, for non-exixtence of four other components, as
\begin{widetext}
\begin{equation}\label{20}
\psi_{0}^{A}(x)=\psi_{0}^{C}(x)=\psi_{0}^{E}(x)=\psi_{0}^{G}(x)=N_{1}Exp\Bigg[-\frac{\Delta_{0}^2\Big(\frac{-\xi Arc\tanh(\tanh(\frac{a(x_{0}-x)}{\xi}))}{a}+\frac{\xi \tanh(\frac{a(x_{0}-x)}{\xi})}{a}+\frac{\xi \tanh^3(\frac{a(x_{0}-x)}{\xi})}{3a}\Big)}{v}\Bigg]
\begin{bmatrix} 1\\0 \end{bmatrix}
\tag{A6}\end{equation}
%\end{widetext}
Moreover, for $\psi_{0}^{A}(x)=\psi_{0}^{C}(x)=\psi_{0}^{E}(x)=\psi_{0}^{G}(x)=0$, the other four solution calculated as
%\begin{widetext}
\begin{equation}\label{21}
\psi_{0}^{B}(x)=\psi_{0}^{D}=\psi_{0}^{F}(x)=\psi_{0}^{H}(x)(x)=N_{2}Exp\Bigg[\frac{\Delta_{0}^2\Big(\frac{-\xi Arc\tanh(\tanh(\frac{a(x_{0}-x)}{\xi}))}{a}+\frac{\xi \tanh(\frac{a(x_{0}-x)}{\xi})}{a}+\frac{\xi \tanh^3(\frac{a(x_{0}-x)}{\xi})}{3a}\Big)}{v}\Bigg]
\begin{bmatrix} 0\\1 \end{bmatrix}
\tag{A7}\end{equation} 
\end{widetext}
in which $N_{1}$ and $N_{2}$ are the normalization factors and $v=\frac{8}{\hbar}(\Delta^2t^2-t^4-\Delta^4/4)$ refers a velocity here for $\theta=\pi/4$. The Eqs.(\ref{20}) and (\ref{21}) indicate that the first, third, fifth, and seventh components of the wave function will be completely localized at the left but on $A$, $C$, $E$, and $G$ sites respectively while the other components show localization on the rest four alternative sites of the chain. A graphical illustration of the same is reported in Fig.\ref{piby4}.

\begin{thebibliography}{00}
\bibitem{comment}The sign of $\Delta$ is insignificant for the energy spectrum while the sign changing, and the related band inversion is considerable for the wave functions, and the topological properties\cite{dw}.
\bibitem{comment1}The consideration of different choices, e.g. $k\rightarrow\pi/a + \hat{p}/\hbar$ has no significant effect on the continuum Hamiltonian resulting in ZESs with the same energy for the same profile and no rotation found for the wave functions.
\end{thebibliography}
\end{document}